# Photon-Counting CT in Cancer Radiotherapy: Technological Advances and Clinical Benefits

Keyur D. Shah, PhD, Jun Zhou, PhD, Justin Roper, PhD, Anees Dhabaan, PhD,
Hania Al-Hallaq, PhD, Amir Pourmorteza, PhD and Xiaofeng Yang, PhD*

[1]Department of Radiation Oncology and Winship Cancer Institute, Emory University, Atlanta, GA 30322
[2]Department of Radiology and Imaging Sciences and Winship Cancer Institute, Emory University, Atlanta, GA 30322
*Email: xiaofeng.yang@emory.edu

**Running title**: Photon-Counting CT in Cancer Radiotherapy

**Manuscript Type:** Review Manuscript

## Abstract

Photon-counting computed tomography (PCCT) marks a significant advancement over conventional energy-integrating detector (EID) CT systems. This review highlights PCCT's superior spatial and contrast resolution, reduced radiation dose, and multi-energy imaging capabilities, which address key challenges in radiotherapy, such as accurate tumor delineation, precise dose calculation, and treatment response monitoring. PCCT’s improved anatomical clarity enhances tumor targeting while minimizing damage to surrounding healthy tissues. Additionally, metal artifact reduction (MAR) and quantitative imaging capabilities optimize workflows, enabling adaptive radiotherapy and radiomics-driven personalized treatment. Emerging clinical applications in brachytherapy and radiopharmaceutical therapy (RPT) show promising outcomes, although challenges like high costs and limited software integration remain. With advancements in artificial intelligence (AI) and dedicated radiotherapy packages, PCCT is poised to transform precision, safety, and efficacy in cancer radiotherapy, marking it as a pivotal technology for future clinical practice.

**Abbreviations:**

- PCCT: Photon-counting CT
- EID: Energy-integrating detector
- OAR: Organs-at-risk
- IMRT: Intensity-modulated radiation therapy
- SBRT: Stereotactic body radiation therapy
- SPR: Stopping power ratio
- MAR: Metal artifact reduction
- ART: Adaptive radiotherapy
- VMIs: Virtual monoenergetic images
- DECT: Dual-energy CT
- MECT: Multi-energy CT
- ULD: Ultra-low dose
- CNR: Contrast-to-noise ratio
- SNR: Signal-to-noise ratio
- CTDIvol: Computed Tomography Dose Index Volume

# 1. Introduction

Computed Tomography (CT) imaging has revolutionized the field of medical diagnostics since its introduction in the 1970s (McCollough 2019). By combining multiple x-ray projections taken from different angles, CT creates detailed cross-sectional images of the body. These images provide crucial insights into the anatomical structures and pathological conditions, aiding in the diagnosis and management of various diseases. The high-resolution images generated by CT scans allow for precise visualization of bones, soft tissues, and blood vessels, making it an indispensable tool in modern medicine (Grüneboom *et al* 2019).

The advancement of CT technology over the years has significantly improved image quality, reduced scan times and minimized radiation exposure. Innovations such as helical CT, multi-detector CT, and iterative reconstruction techniques have enhanced the capabilities of CT imaging, enabling faster and more accurate diagnostics (Hsieh and Flohr 2021). These developments have expanded the applications of CT beyond traditional diagnostic purposes, paving the way for its integration into therapeutic procedures (Goitein *et al* 1979), particularly in cancer treatment workflows like radiotherapy.

Radiotherapy, a cornerstone of cancer treatment, involves the use of high-energy radiation to destroy cancer cells while sparing healthy tissues. The success of radiotherapy heavily relies on precise targeting and accurate dose delivery, which are facilitated by high-quality imaging. CT scans are integral throughout the radiotherapy workflow, starting from the initial diagnosis and tumor localization to treatment planning and monitoring. By providing detailed anatomical information, CT scans enable clinicians to delineate tumors and organs-at-risk (OARs) accurately, which is crucial for developing effective treatment plans. Additionally, CT images are used to create three-dimensional (3D) models of the patient's anatomy, allowing for precise dose calculations and optimization of radiation delivery to target the cancer while minimizing dose to healthy nearby organs.

Despite its critical role, conventional CT imaging has limitations that can impact the accuracy and effectiveness of radiotherapy. These include suboptimal contrast resolution, artifacts, and the inability to provide functional information (Pereira *et al* 2014). Moreover, conventional CT systems struggle with accurate differentiation between tissues of similar density, which is particularly important for radiotherapy delivered by proton beams, and are prone to beam-hardening artifacts, which can compromise image quality. Addressing these challenges is essential to further enhancing the precision and outcomes of radiotherapy.

Photon-Counting CT (PCCT) offers promising solutions to overcome the limitations of conventional CT and significantly improve the radiotherapy process. PCCT represents a paradigm shift in medical imaging, particularly in the context of cancer treatment. By offering enhanced spatial and contrast resolution, reduced artifacts, and the ability to perform multi-energy imaging (Willemink *et al* 2018, Flohr *et al* 2020a), PCCT can produce advanced images, such as virtual monoenergetic images (VMIs), iodine maps, and virtual non-contrast (VNC) images. These capabilities are particularly useful for radiotherapy, where precise tumor delineation and dose calculations are critical. Iodine maps, for example, can act as surrogates for tumor perfusion, allowing clinicians to monitor treatment response and track changes in tumor vascularization during radiotherapy. Furthermore, the ability of PCCT to generate data on relative electron density and effective atomic number directly supports more accurate dose calculations, improving treatment planning and overall outcomes. The integration of PCCT into clinical practice promises to advance the precision and efficacy of cancer treatments, marking a significant step forward in personalized medicine.

While significant advancements have been made in CT technology, reviews discussing PCCT often focus on its applications in radiology and diagnostic medicine. There is a noticeable gap in the literature examining its potential within the radiotherapy context. Existing reviews and studies tend to emphasize technological innovations of PCCT or Dual-Energy CT (DECT) (Willemink *et al* 2018, Jacobsen *et al* 2020, Flohr *et al* 2020a, Si-Mohamed *et al* 2021, Farhadi *et al* 2021, Kruis 2022, Douek *et al* 2023, Wehrse *et al* 2023, van der Bie *et al* 2023, Meloni *et al* 2024). However, there is a lack of focused analysis linking these technological advancements to practical applications in radiotherapy. This review aims to fill this void by offering a detailed exploration of how PCCT can enhance the precision and efficacy of radiotherapy, thereby providing a valuable resource for clinicians, researchers, and physicists seeking to optimize cancer treatment outcomes.

Figure 1 below illustrates the growing body of literature related to PCCT in recent years, showing its increasing popularity across various medical applications. However, despite this growth, its application in radiotherapy remains significantly underexplored. Given PCCT's potential to address key challenges in radiotherapy—such as improving tumor targeting and dose accuracy—this gap presents a critical opportunity for further research and clinical integration. The rising interest in PCCT across other medical disciplines signals its transformative potential in radiotherapy, making it an essential area for future exploration.

The rest of the review paper is organized as follows: Section 2 outlines the methodology used for data collection. Section 3 provides an overview of the fundamental physics behind photon-counting detectors. In Section 4, we delve into the advancements in spectral CT technology. Section 5 highlights the current state of commercially available PCCT systems. Section 6 explores how PCCT can enhance the radiotherapy workflow. Sections 7 and 8 focus on the transformative potential of PCCT in deep learning applications and radiopharmaceutical therapy, respectively. Section 9 addresses the impact of PCCT on reducing radiation dose in CT scans. Finally, Section 10 summarizes the key contributions of this review and discusses the limitations associated with PCCT technology.

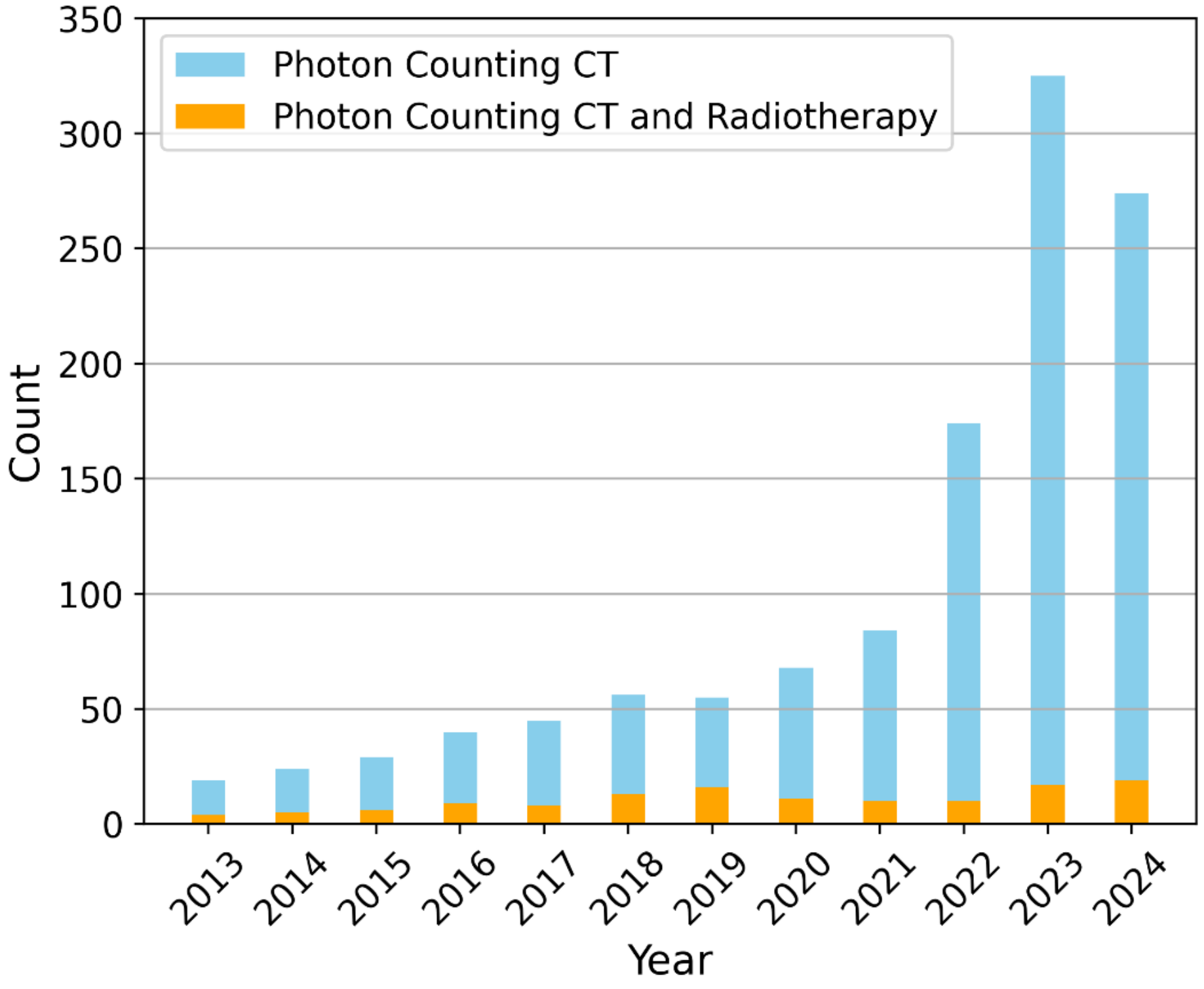

**Figure 1.** Publication trends in Photon-Counting CT (PCCT) and its application in radiotherapy from 2013 to 2024*. The figure shows a steep rise in the number of studies focusing on PCCT in general, but a relatively low number of studies specifically addressing its use in radiotherapy, emphasizing the need for further exploration in this area.

* Data for 2024 includes publications until November 25, 2024. (Source: PubMed)

## 2. Data Collection

To ensure a comprehensive and systematic review of the literature on PCCT and its applications in radiotherapy, we employed a structured data collection approach. Our methodology involved the following steps:

**Database Search:** We utilized PubMed as our primary database for sourcing relevant articles. The search query used was "(Photon counting OR Spectral Photon counting) AND CT) AND (radiotherapy OR radiation therapy)". The search was conducted for articles published up to November 25th, 2024.

**Filtering Process:** The initial search yielded a total of 162 articles. To refine the results, we applied the following filters:

1. Publication type: Peer-reviewed articles

- Language: English
- Relevance: Articles that specifically address the use of PCCT in the context of radiotherapy

After applying these filters, we obtained a subset of 103 articles.

**Independent Citation Analysis:** To ensure the inclusion of highly relevant and influential studies, we conducted an independent citation analysis. This process involved reviewing the reference lists and citations of the initially selected articles. By examining the papers cited frequently by our selected articles, we identified additional key studies that contributed significantly to the field. This method allowed us to uncover influential research that may not have appeared in the initial search results.

**Final Selection:** The final selection comprised 116 articles, which were thoroughly reviewed and analyzed to extract relevant data and insights on the advancements, applications, and impact of PCCT in radiotherapy. These articles provided a comprehensive understanding of the current state of research and potential future directions in this domain.

**Limitations of the Search:** While PubMed is a comprehensive database, it may not capture all relevant articles, especially those published in less mainstream journals or in languages other than English or manuscripts published on arXiv. Additionally, the focus on English-language articles may have introduced a language bias, potentially excluding relevant studies published in other languages.

The complete data collection process is summarized in Figure 2, which presents a PRISMA diagram outlining the search, filtering, and selection of studies included in this review.

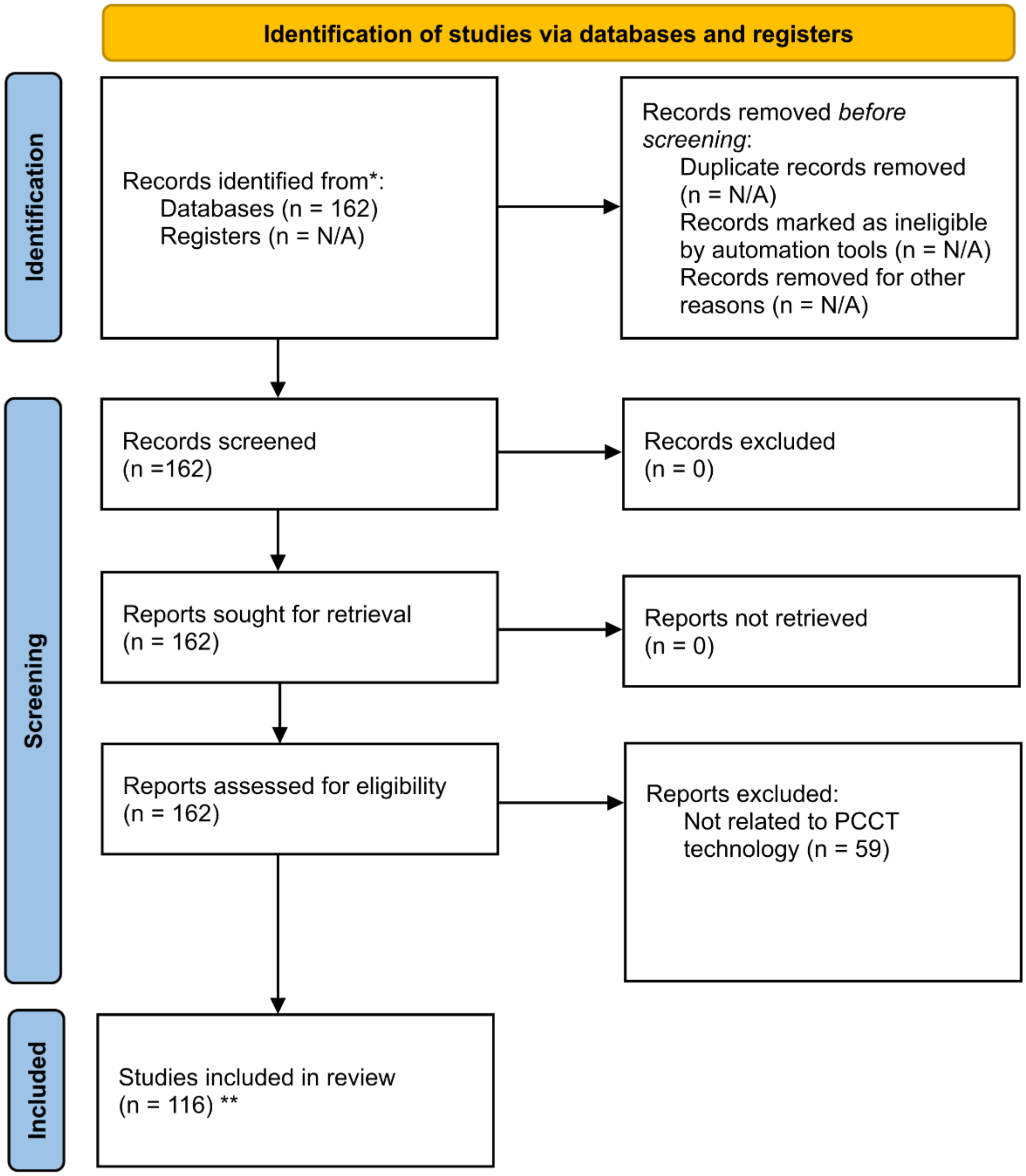

**Figure 2**. PRISMA flow diagram outlining the systematic search, screening, and selection process for studies included in the review on Photon-Counting CT (PCCT) and its applications in radiotherapy. The diagram highlights the identification of relevant studies from databases, the filtering process, reasons for exclusion, and the final number of studies included in the review.(Page et al 2021). ** n = 13 studies were added after independent citation analysis.

## 3. Physics of Photon-Counting Detectors vs. Energy-Integrating Detectors

To understand the advancements brought by PCCT, it is essential to delve into the fundamental differences between photon-counting detectors (PCDs) and conventional energy-integrating detectors (EIDs).

**Energy-Integrating Detectors (EIDs):** Traditional CT systems are energy-integrating (i.e., EICT), which measure the total energy deposited by x-ray photons. When the detector material absorbs incoming x rays, the resulting electrical signal is proportional to the total energy deposited. This signal is then integrated over a set time period to form the image (Fig. 3 A). The materials used in EIDs typically include scintillators and photodiodes. Common scintillator materials are made of materials like cesium iodide (CsI) or gadolinium oxysulphide (GOS), which convert x-ray photons into visible light. The visible light is then detected by photodiodes, such as amorphous silicon (a-Si) photodiodes or photomultiplier tubes (PMTs), which convert the light into an electrical signal. This signal is collected and processed to create the final CT image. The choice of scintillator material affects the detector's efficiency, spatial resolution, and sensitivity to x rays. The pixel sizes in EIDs generally range from 0.5 mm to 1.0 mm, which limits spatial resolution, especially when capturing fine details.

**Limitations of EIDs:**

1. **Noise and Resolution:** EIDs are susceptible to electronic noise, which can degrade image quality. They also have limited spatial resolution because the scintillator crystals and photodiodes in the detector are relatively large, and light crosstalk between pixels must be minimized using reflecting septa. This design inherently limits the detector's ability to capture fine spatial details.
2. **Energy Discrimination:** EIDs do not differentiate between photons of different energies. This lack of energy discrimination can lead to suboptimal contrast resolution and reduced ability to distinguish between different tissue types.
3. **Beam Hardening Artifacts:** EIDs are prone to beam hardening artifacts, where lower energy photons are preferentially attenuated as compared to higher energy photons, leading to undesirable contrast variations and streaking artifacts in the image.

**Photon-Counting Detectors (PCDs):** In contrast, PCDs represent a significant technological advancement because they directly count individual photons and measure their energy. When an x-ray photon interacts with the detector, it generates an electrical pulse whose amplitude is proportional to the photon's energy. These pulses are counted and categorized based on their energy levels (Fig. 3 B). The materials used in PCDs are crucial for their performance and typically include semiconductor materials like cadmium telluride (CdTe) or cadmium zinc telluride (CZT). These materials have high atomic numbers and are effective at converting x-ray photons into electronic signals. In a PCD, the semiconductor material is usually configured into an array of small pixels, each capable of detecting x-ray photons and measuring their energy with high precision. The pixel sizes in PCDs are much smaller, typically ranging from 0.1 mm to 0.2 mm, allowing for significantly improved spatial resolution. The signal generated by each pixel is then processed to produce detailed images with enhanced contrast and spatial resolution. The use of these advanced materials allows PCDs to achieve superior energy resolution and higher image quality compared to traditional EIDs.

**Advantages of PCDs:**

1. **Ultra-High Spatial Resolution:** PCDs offer higher spatial resolution compared to EIDs, allowing for more detailed and precise imaging PCDs offer higher spatial resolution compared to EIDs, allowing for more detailed and precise imaging because each pixel in the detector is smaller and more sensitive to individual photons. This fine pixelation, combined with the ability to count and categorize photons directly, results in sharper images with improved edge definition, which is crucial for detecting small lesions or fine anatomical details.
2. **Superior Contrast Resolution:** By discriminating between photons of different energies, PCDs provide superior contrast resolution, enabling better differentiation between tissues.
3. **Reduced Noise:** PCDs inherently reduce electronic noise because they count individual photons, resulting in clearer and more accurate images.
4. **Multi-Energy Imaging:** PCDs can perform multi-energy imaging in a single scan, providing rich spectral information that can be used to better characterize tissue composition.
5. **Artifact Reduction:** PCDs help reduce beam hardening artifacts by enabling more precise energy discrimination. By counting and categorizing photons at different energy levels, PCDs allow for better differentiation of materials, minimizing the appearance of beam hardening artifacts and resulting in clearer, more reliable images.

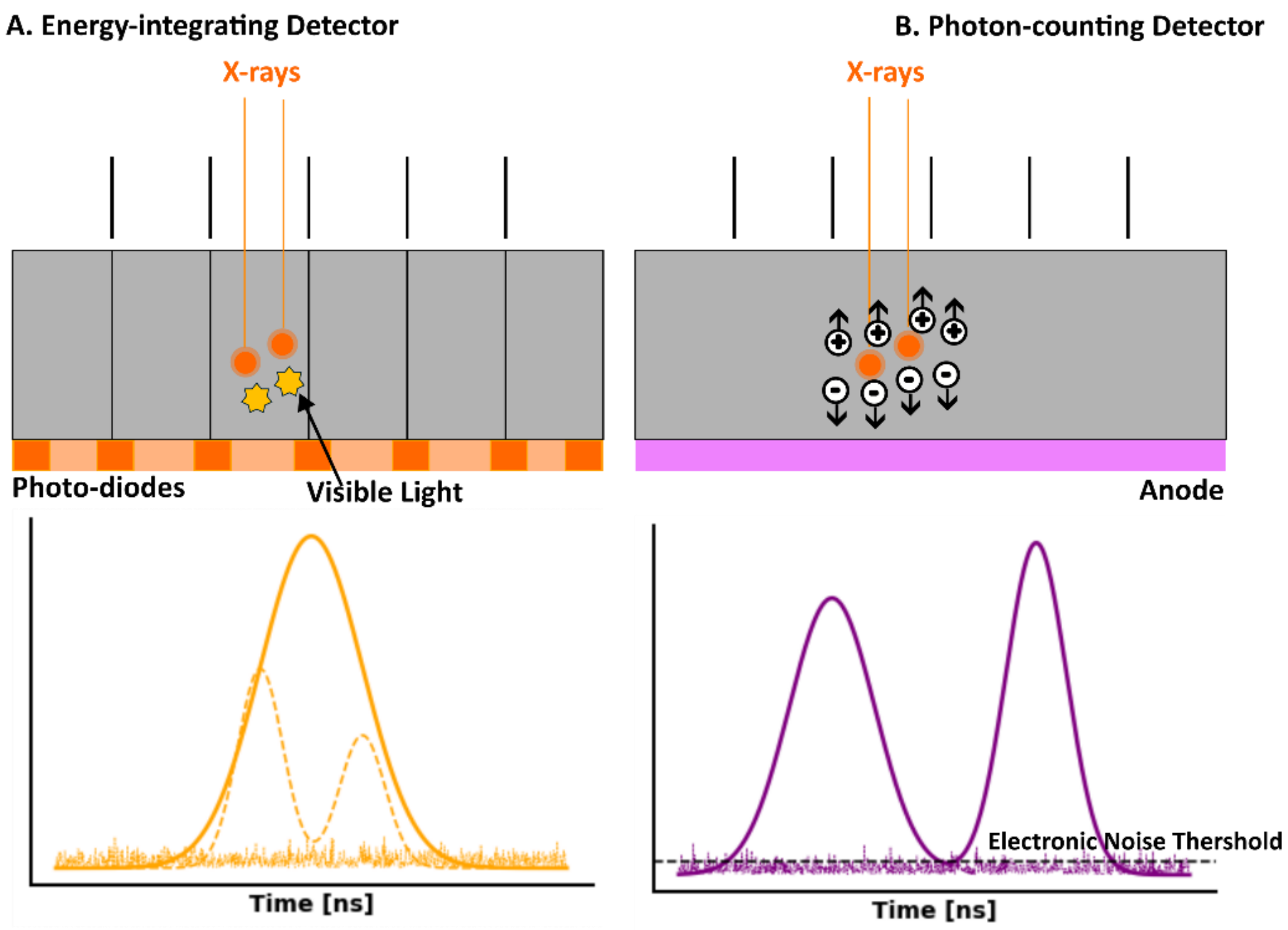

**Figure 3. Schematic Overview of CT Detectors: A.** Energy-integrating detectors convert incoming x-ray photons into visible light, which is then transformed into an electrical signal by photodiodes. The signal for each projection is obtained by integrating all detected photons within that projection. The spatial resolution of these detectors is influenced by the use of reflecting septa to minimize crosstalk between detector elements. **B.** Photon-counting detectors generate an electron-hole pair when a photon interacts with the semiconductor material. The anode attracts the electrons from this pair, producing an electrical pulse whose amplitude corresponds to the energy of the incoming photon. Unlike EIDs, where pixel size and photon integration over time limit spatial resolution, PCDs offer continuous photon sampling and count individual photons, resulting in higher spatial resolution and enhanced image quality. Adapted from (van der Bie et al 2023)

Rajendran et al (Rajendran *et al* 2022) demonstrated that PCCT offers significant advantages over traditional EICT in coronary CT angiography. In their study involving a 71-year-old man (Fig. 4), PCCT enabled multi-energy imaging with 66-msec temporal resolution, which is not possible with EICT. The 45- and 55-keV VMIs generated by the PCCT system exhibited higher iodine signal levels (1164 HU at 45 keV and 800 HU at 55 keV) compared to the 90 kV EID CT images (724 HU), despite the PCCT using 22% less iodine contrast material (90 mL vs. 110 mL). Additionally, the PCCT iodine maps provided clear visualization of the left coronary artery without motion blur, a significant improvement over EICT. The inability of EICT to create VMIs, iodine maps, and VNC images at this temporal resolution highlights the superior capabilities of PCCT in cardiac imaging, particularly for applications requiring detailed multienergy data. The VNC image, shown in the bottom row, is generated by subtracting iodine from contrast-enhanced data to simulate a non-contrast image without requiring a separate scan. In this case, the VNC image shows 72 HU in the region of interest, closely resembling traditional non-contrast images but derived from the same contrast-enhanced dataset.

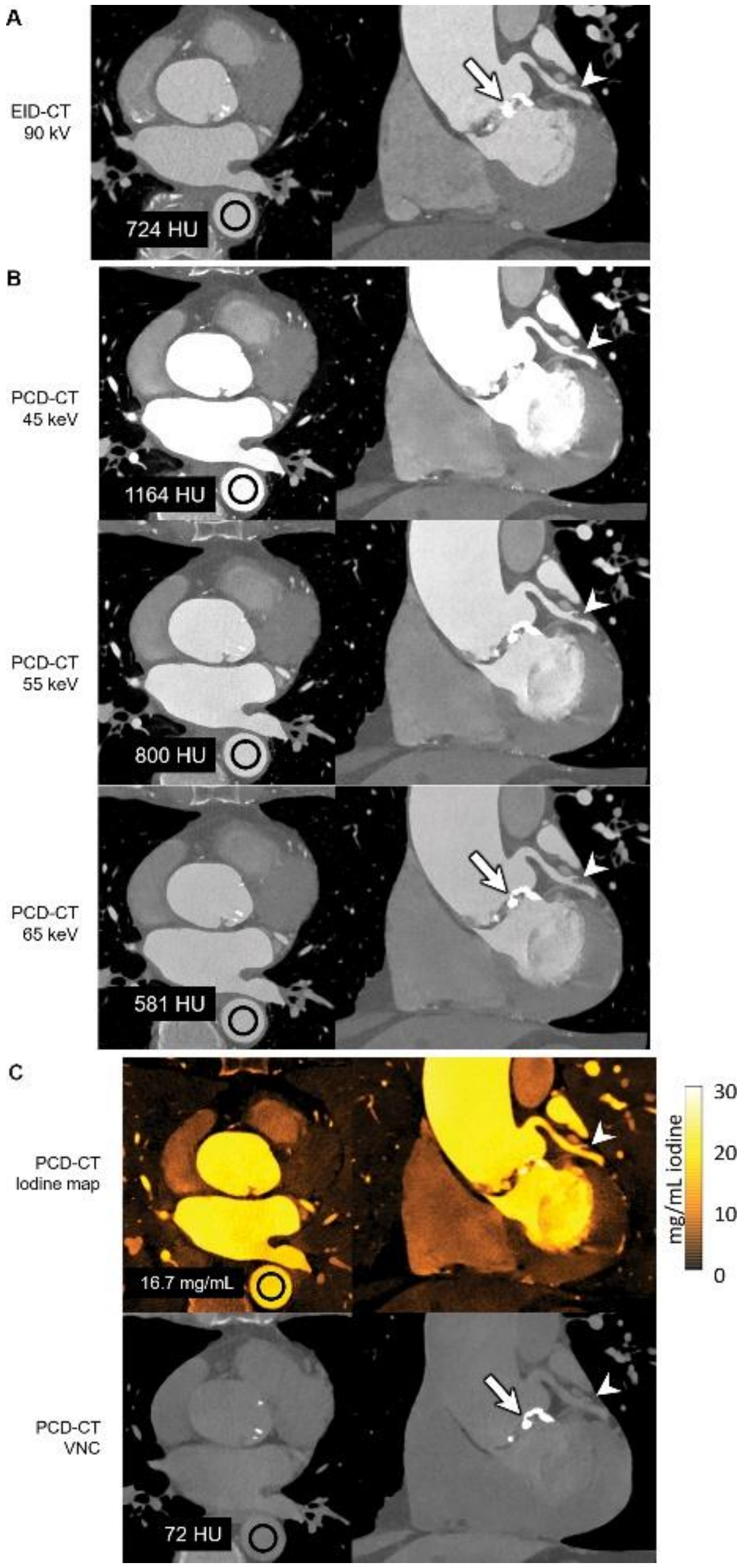

**Figure 4**. Images in a 71-year-old man scanned with (A) energy-integrating detector (EID) based CT and (B, C) PCCT. The multienergy capabilities of the PCCT system allowed the creation of low-energy virtual monoenergetic images (VMIs) (B), showing increased iodine signal and better delineation of coronary arteries compared to single-energy

EICT (A). The bottom row (C) also includes a Virtual Non-Contrast (VNC) image, created by subtracting iodine from the contrast-enhanced dataset, which effectively mimics a non-contrast image. Reconstruction kernels and display settings varied between EICT and PCCT. This figure has been adapted from the work of Rajendran et al (Rajendran et al 2022) with permission from the journal.

Rajagopal et al (Rajagopal *et al* 2021) conducted a study comparing the quantitative image quality of PCCT to traditional EICT across various low-dose levels in phantoms. Using an investigational scanner equipped with both PCCT and EICT subsystems, they evaluated image quality at dose levels of 1.7, 2, 4, and 6 mGy CT dose index volume (CTDIvol), all of which are at or below the doses typically used for conventional abdominal CT. The results demonstrated that PCCT significantly outperformed EICT in terms of image quality. Specifically, PCCT images showed a 22.1%−24.0% reduction in noise across the dose levels, leading to a 29–41% improvement in contrast-to-noise ratio (CNR) and a 20–36% enhancement in detectability index. Furthermore, for iodine detection, PCCT consistently provided higher CNR across all doses and iodine concentrations evaluated.

Similarly, Stein et al (Stein *et al* 2023) investigated the impact of PCCT on small vessel stent visualization compared to traditional EICT. Their study found that PCCT with a dedicated sharp vascular kernel (Bv56) provided superior image quality compared to EICT in phantom. The highest diagnostic confidence was observed with PCCT, particularly in terms of sharpness and reduced blooming artifacts, which are critical for accurate stent assessment. The study also demonstrated that PCCT could potentially reduce the need for invasive coronary angiograms by improving non-invasive imaging quality.

The fundamental differences between EIDs and PCDs underscore the transformative potential of PCCT in radiotherapy. By overcoming the limitations of traditional EIDs—such as noise, resolution, and artifact issues—PCDs pave the way for enhanced imaging capabilities that are particularly beneficial in the precise delivery of treatments.

## 4. Spectral CT: Dual-Energy CT (DECT) vs. Photon-Counting CT (PCCT)

Multi-energy CT (MECT), as described by Rajiah et al (Rajiah *et al* 2020), involves acquiring two or more CT measurements with distinct energy spectra, enabling a more detailed differentiation of tissues and materials than conventional CT. This advanced imaging technique leverages the energy-dependent attenuation properties of x-ray photons to enhance tissue characterization and improve diagnostic accuracy. The two primary modalities of Spectral CT are Dual-Energy CT (DECT) and PCCT, each utilizing distinct technological approaches to exploit spectral information. These modalities allow for advanced imaging techniques such as K-edge imaging, Z_eff mapping, VNC imaging, and VMI.

Sauerbeck et al (Sauerbeck *et al* 2023) in their review, demonstrated that Spectral CT's advanced imaging capabilities, especially in oncology, offer significant advantages for radiotherapy planning and monitoring. By generating virtual unenhanced images, iodine maps, and VMIs, Spectral CT enables more precise detection and characterization of tumors. In the context of radiotherapy, these capabilities can be used to improve the accuracy of tumor delineation and treatment response assessment. For example, iodine maps, which act as a surrogate for tumor perfusion, can be utilized to monitor the effectiveness of radiotherapy by tracking changes in tumor vascularization. Additionally, the ability to calculate relative electron density and the effective atomic number from Spectral CT data is crucial for accurate dose calculation and optimization in radiotherapy planning.

**Dual-Energy CT (DECT)**

**Principles**: DECT operates by acquiring CT images at two different energy levels, typically using either two distinct x-ray sources or by rapidly switching between high and low-energy x rays during a single scan. This dual-energy data enables differentiation of tissues based on their unique energy-dependent attenuation characteristics, making it possible to achieve material decomposition and enhanced contrast resolution.

**Applications and Advantages**:

- **Material Differentiation**: DECT is particularly effective in distinguishing between different types of tissues and materials. For example, it can differentiate iodine from calcium, which is valuable in vascular imaging and in identifying calcifications. In radiotherapy, this capability can improve the accuracy of tumor delineation by better distinguishing tumors from surrounding healthy tissues or calcifications, which is critical for precise dose delivery.

- **Enhanced Contrast Resolution**: By analyzing attenuation at two different energy levels, DECT enhances contrast resolution, making it easier to visualize structures such as blood vessels, tumors, and other soft tissues. For radiotherapy, this improved contrast can lead to more accurate identification of the tumor boundaries and organs-at-risk (OARs), optimizing treatment planning and minimizing radiation exposure to healthy tissues.
- **Artifact Reduction**: DECT can help reduce beam-hardening artifacts and other common CT artifacts, leading to improved image quality. This is particularly important in radiotherapy, where artifact-free images are essential for accurate tumor localization and dose calculation.
- **Functional Imaging**: DECT enables functional imaging, such as iodine mapping in perfusion studies, which provides additional diagnostic information on tissue perfusion and vascularity. In radiotherapy, iodine maps can serve as a surrogate for tumor perfusion, allowing for dynamic monitoring of treatment response and adaptation of the treatment plan based on changes in tumor vascularity.

Recent advancements further extend DECT's capabilities. Peng et al (Peng *et al* 2024) introduced an unsupervised-learning framework for material decomposition in DECT, addressing one of its biggest challenges: noise amplification during material decomposition. This deep-learning-based model demonstrated significant noise reduction (up to 97%) without requiring paired data, enhancing DECT's potential for quantitative imaging in radiotherapy applications. Similarly, Gao et al (Gao *et al* 2024) employed a Conditional Denoising Diffusion Probabilistic Model (C-DDPM) to generate synthetic contrast-enhanced DECT (CE-DECT) images from non-contrast single-energy CT (SECT) scans. This method is especially useful for patients who are at risk from iodinated contrast agents, and for institutions lacking DECT scanners, offering a novel solution for radiation therapy planning with minimal imaging risks.

**Photon-Counting CT (PCCT)**

**Principles**: PCCT employs PCDs that count individual x-ray photons and measure their energy directly. This direct measurement allows for precise energy discrimination, high-resolution imaging, and the acquisition of multi-energy data in a single scan.

**Applications and Advantages**:

- **Ultra-High Spatial Resolution**: PCCT offers superior spatial resolution by counting individual photons and minimizing electronic noise, which improves the clarity of images. In radiotherapy, this high spatial resolution is critical for accurately delineating small tumors and detailed anatomical structures, ensuring precise dose delivery and minimizing radiation exposure to surrounding healthy tissues.
- **Superior Contrast Resolution**: The ability of PCCT to perform direct energy discrimination leads to improved contrast resolution, allowing for finer differentiation between tissues. This is particularly valuable in radiotherapy for distinguishing between tumors and nearby organs-at-risk (OARs), helping to define treatment volumes more precisely and improving the safety and efficacy of treatment planning.
- **Multi-Energy Imaging**: Unlike DECT, which is limited to two energy levels, PCCT can acquire data across multiple energy bins simultaneously, providing more comprehensive spectral information. In radiotherapy, this multi-energy capability can be used for advanced tissue characterization, enabling more precise adjustments to treatment plans based on the varying tissue properties and improving the accuracy of dose calculations.
- **Artifact Reduction**: PCCT significantly reduces common artifacts, such as beam hardening, resulting in clearer and more accurate images. In radiotherapy, artifact-free images are crucial for precise tumor localization, ensuring that the treatment targets only cancerous tissue and avoids damage to healthy structures.
- **Quantitative Imaging**: PCCT enables precise quantitative imaging, such as calculating electron densities and tissue compositions, which is highly useful for dosimetry in radiotherapy. This quantitative data enhances the accuracy of treatment planning by allowing more refined calculations of radiation dose distributions, leading to improved treatment outcomes.

A recent study by Ren et al (Ren *et al* 2024) evaluated the spectral imaging performance of a clinical PCCT system for single- and dual-contrast materials in comparison to DS-DECT. The study showed that while PCCT provided useful spectral imaging capabilities, DS-DECT with 70/Sn150 kV or 80/Sn150 kV offered superior accuracy in two-material decomposition tasks. For instance, root-mean-square-error (RMSE) values for iodine and gadolinium were lower in DS-DECT compared to PCCT, especially in dual-contrast tasks. This highlights that while PCCT holds great promise, it may still be outperformed by advanced dual-energy techniques like DS-DECT for certain clinical applications. The greater spectral separation in DS-DECT's energy levels likely contributed to its improved material decomposition performance, particularly in complex imaging scenarios.

Moreover, Deng et al (Deng *et al* 2024) explored the integration of PCCT in dual-contrast imaging, demonstrating the ability to differentiate between iodine and gadolinium in clinical scenarios. Their findings revealed that PCCT is particularly advantageous in three-material decomposition, where RMSE values were within clinically acceptable ranges for complex tissue compositions. This capability underscores the potential of PCCT to advance multi-contrast imaging, particularly in oncology where precise material differentiation may be vital for tumor delineation and therapy monitoring.

Table 1 compares the key features and advantages of DECT and PCCT, illustrating the technological differences in terms of energy acquisition, spatial and contrast resolution, artifact reduction, and quantitative imaging capabilities. Figure 5 further illustrates these advancements by comparing VNC images from PCCT and EICT. The PCCT image shows reduced noise, better delineation of small structures, and overall enhanced image quality, which is critical for applications like tumor targeting and segmentation in radiotherapy.

**Table 1.** Comparison: DECT vs. PCCT.

| **Feature** | **Dual-Energy CT** | **Photon-Counting CT** |
|---|---|---|
| Energy Acquisition | Two energy levels (dual-source or rapid kV switching) | Multiple energy bins (photon-counting detectors) |
| Spatial Resolution | Moderate, limited by noise | High, reduced electronic noise |
| Contrast Resolution | Enhanced with dual-energy data | Superior with direct energy discrimination |
| Artifact Reduction | Beam hardening reduction | Significant reduction, especially beam hardening |
| Quantitative Imaging | Limited, depends on dual-energy data | Precise, suitable for dosimetry |
| Complexity | More complex, requires dual-source or kV switching | Simpler in concept, but technology still maturing |
| Cost | Expensive due to dual-source technology | Higher due to advanced detector technology |

## Advanced Spectral CT Concepts

- **Z_eff Imaging**: The effective atomic number (Z_eff) of tissues can be calculated using DECT and PCCT, aiding in tissue characterization and material differentiation. This is crucial in distinguishing between materials with similar attenuation coefficients at specific energies (Landry *et al* 2013). PCCT's ability to acquire data across multiple energy bins may further refines Z_eff imaging, potentially contributing to more accurate treatment planning and outcome prediction in radiotherapy. This technique is especially valuable in Stereotactic Radiosurgery (SRS) cases, where patients may not be eligible for magnetic resonance imaging (MRI) and CT with iodine contrast is used instead. Z_eff imaging can enhance tumor delineation by better characterizing iodine, which crosses the blood-brain barrier and localizes in small tumors. Given the tight contour margins and the risk of radionecrosis, precise target delineation is critical for effective treatment. By improving the contrast of iodine uptake, PCCT Z_eff imaging offers a more accurate alternative in such cases, where CT contrast resolution is typically poorer compared to MRI.
- **Virtual Non-Contrast (VNC) Imaging**: VNC imaging subtracts the iodine component from contrast-enhanced images, providing an image similar to a non-contrast scan (Mergen *et al* 2022). This technique eliminates the need for a separate non-contrast acquisition, reducing the patient's radiation exposure and the risks associated with repeated administration of contrast agents. VNC imaging is particularly useful in radiotherapy, where multiple imaging sessions may be required. For example, in Liver SBRT cases, where CT contrast between the tumor and surrounding liver tissue is often poor, VNC imaging can improve the visualization of the target without additional contrast exposure, helping clinicians avoid repeat contrast-enhanced scans.
- **Virtual Monoenergetic Imaging (VMI)**: VMI generates images at single energy levels, offering improved contrast at specific keV levels (Li *et al* 2021). This technique is available in both DECT and PCCT and is beneficial in enhancing contrast resolution and reducing artifacts. In radiotherapy, VMI can be used to optimize image contrast for better visualization of tumors and surrounding tissues, aiding in precise targeting and dose delivery. For example, a study conducted on monoenergetic CT images from dual-energy scanning by (Pawałowski *et al* 2019) demonstrated that images reconstructed at 70 keV offered the best combination of low contrast resolution, noise, and signal-to-noise ratio (SNR). This suggests that 70 keV might be the optimal energy level for delineating structures in radiotherapy planning, due to its superior uniformity and SNR. Incorporating this into the planning process could enhance the accuracy of tumor delineation and improve overall treatment outcomes. Racine et al (Racine *et al* 2024) demonstrated that PCCT's advanced monoenergetic imaging capabilities allow for optimized lesion detectability across varying radiation dose levels and patient sizes. VMIs at 65-70 keV provided the highest detectability, reinforcing PCCT's role in enhancing image quality and supporting quantitative radiomics analyses for precise tumor characterization.

Dane et al (Dane *et al* 2024b) conducted a study comparing the image quality of portal venous phase-derived VNC images obtained from PCCT and energy-integrating dual-energy computed tomography (EI-DECT). The study involved a cohort of 74 patients, where both qualitative and quantitative analyses were performed to assess the image quality differences between the two modalities. The findings demonstrated that PCCT VNC images consistently provided superior overall image quality, reduced noise, and better delineation of small structures compared to EI-DECT. Additionally, PCCT exhibited improved CNR and SNR, especially in non-enhancing structures like fat, making it particularly advantageous in clinical scenarios where precise tissue characterization is critical. Notably, PCCT achieved these improvements with comparable radiation doses to EI-DECT, highlighting its potential for enhanced imaging without additional radiation risk. These findings underscore the significant advantages of PCCT over EI-DECT, particularly in achieving higher image quality with reduced artifacts, which is crucial for applications such as radiotherapy planning where precision is paramount.

These findings align with those of Onishi et al (Onishi *et al* 2024), who emphasized that PCCT’s direct conversion of x rays into electrical signals enables it to generate VMIs with high CNR, improving the ability to visualize small or low-contrast lesions. Onishi et al. also noted that PCCT offers significant dose efficiency, with studies showing a dose reduction of approximately 32% in contrast-enhanced abdominal CT while maintaining image quality comparable to second-generation dual-source CT (DSCT). These improvements, including reduced noise and better delineation of structures such as the renal pelvis, ureters, and mesenteric vessels, further underscore PCCT’s advantages in clinical applications requiring high precision, such as radiotherapy planning. Figure 5 presents a sample VNC image from PCCT and EICT.

Similarly, Ren et al (Ren *et al* 2024) conducted a comparative study between PCCT and Dual-Source DECT for multi-contrast imaging using iodine and gadolinium. The study found that DECT with greater spectral separation (such as 70/Sn150 kV and 80/Sn150 kV) outperformed PCCT in terms of reducing root-mean-square-error (RMSE) values for two-material decomposition. However, PCCT demonstrated better spatial resolution and noise reduction in certain imaging tasks, particularly for lower energy imaging and multi-contrast applications.

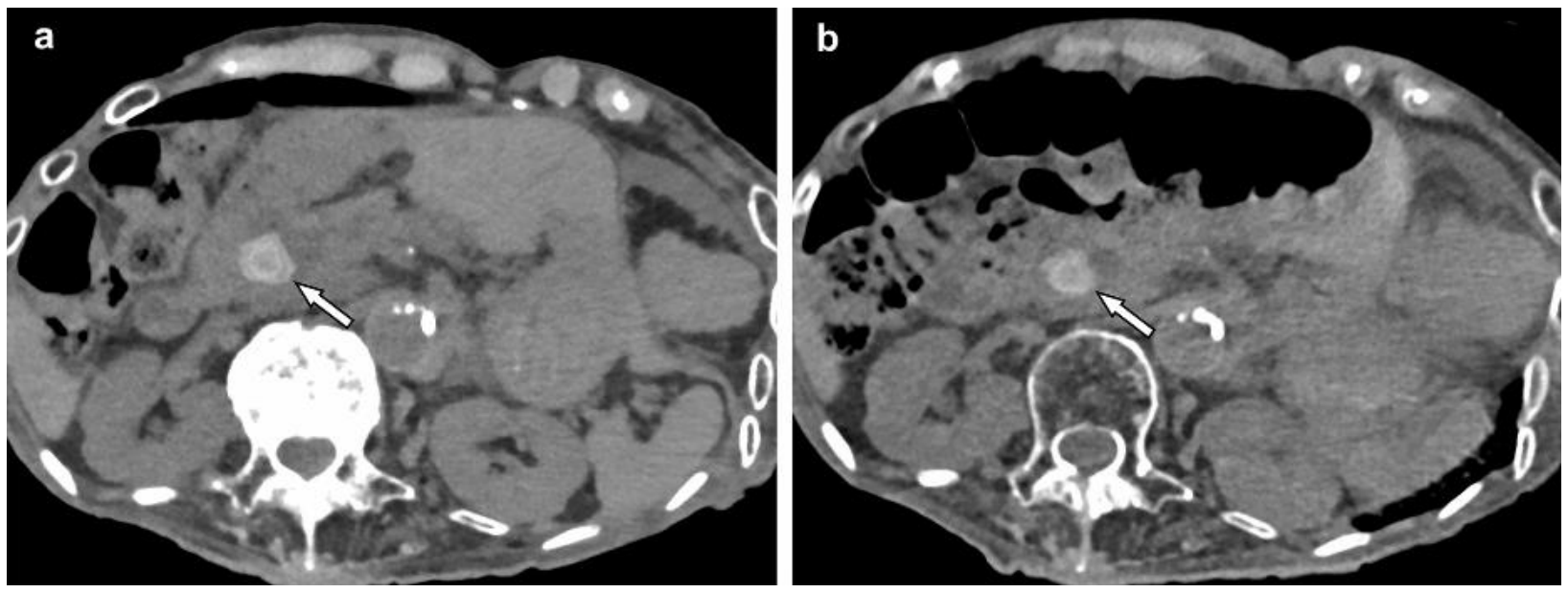

**Figure 5.** Non-contrast abdominal CT images of a patient with choledocholithiasis scanned with PCCT (A) and conventional DSCT in single-energy mode (B), performed on different days. The PCCT image, with a CTDIvol of 7.5 mGy, demonstrates reduced noise, enhanced image quality, and improved visualization of small structures, such as the calcified stone in the common bile duct (arrows), compared to the conventional CT image, which had a CTDIvol of 11.9 mGy. This figure has been adapted from the work of Onishi et al (Onishi et al 2024) with permission from the journal.

## 5. Currently Available PCCT Systems

PCCT technology is undergoing rapid development and is being adopted by several prominent medical institutions worldwide. Leading centers in the United States, such as Duke University, Mayo Clinic, Stanford University, and the National Institutes of Health (NIH), are at the forefront of integrating PCCT into clinical practice and research. These institutions are exploring the benefits of PCCT's improved spatial resolution, spectral imaging capabilities, and radiation dose reduction.

Several manufacturers have developed PCCT systems, with notable differences in detector technology, collimation, scan field of view, and energy thresholds. These manufacturers, including Siemens Healthcare, GE Healthcare, Philips Healthcare, Canon Medical Research,

Samsung/NeuroLogica, and MARS Bioimaging, have each introduced systems with unique specifications tailored to different clinical applications.

Below is a comparison of the currently available PCCT systems, detailing key characteristics such as detector type, collimation, scan field of view, and FDA clearance status. Table 2 summarizes the technology and FDA status of various PCCT systems.

**Table 2.** Characteristics of PCCT systems Modified from (Douek et al 2023).

| **Manufacturer and Product Name** | **Detector Type** | **Collimation** | **No. of Energy Thresholds** | **Body Region** | **Scan FOV (cm)** | **FDA Clearance** |
|---|---|---|---|---|---|---|
| Siemens Healthcare (Naeotom Alpha) | CdTe | 144 x 0.4 mm (SR); 120 x 0.2 mm (UHR) | 4 (SR); 2 (UHR) | Full body | 50 | Yes |
| GE Healthcare | Silicon | * | 8 | Full body | 50 | No |
| Philips Healthcare | CdZnTe | 64 x 0.275 mm | 8 | Full body | 50 | No |
| Canon Medical Research | CdZnTe | 16 x 0.62 mm (SR); 48 x 0.21 mm (UHR) | 6 | Full body | 50 | No |
| Samsung/NeuroLogica (OmniTom Elite PCD) | CdTe | 80 x 0.15 mm | 3 | Head and neck | 25 | Yes |
| MARS Bioimaging (extremity 5 x 120) | CdZnTe | 128 x 0.11 mm | 5 | Extremities | 12, 5 | No |

Note: CdTe = cadmium telluride, CdZnTe = cadmium zinc telluride, FDA = U.S. Food and Drug Administration, FOV = field of view, PCD = photon-counting detector, SR = standard resolution, UHR = ultrahigh resolution. *Collimation data not available for GE Healthcare as it is a prototype.

These systems vary in terms of collimation, energy thresholds, and the regions of the body they are designed to image, offering different levels of resolution and image quality. The Siemens Naeotom Alpha, for instance, is one of the most advanced systems currently available, offering high-resolution imaging for full-body scans and is FDA-cleared for clinical use. Similarly, the Samsung/NeuroLogica OmniTom Elite PCD is FDA approved yet it is designed for head and neck imaging.

The ongoing research and clinical trials conducted at leading institutions are expected to push the boundaries of PCCT's capabilities further, improving diagnostic accuracy and expanding its applications in fields such as oncology, cardiology, and neurology. With continuous technological advancements, PCCT systems are poised to become a cornerstone of advanced medical imaging in the coming years.

## 6. PCCT in Radiotherapy

The introduction of PCDs in CT imaging, leading to the development of PCCT, brings substantial improvements over conventional CT technologies. These advancements address the limitations of EIDs and offer enhanced imaging capabilities that are particularly beneficial for radiotherapy.
In the context of radiotherapy, PCCT's higher spatial and contrast resolution, along with the ability to perform multi-energy imaging, significantly enhances tumor delineation, treatment planning, and dose calculation. These improvements may translate into more precise targeting of cancerous tissues, better sparing of healthy tissues, and overall improved treatment outcomes. In the following sections, we will explore the specific advantages of PCCT in detail and how they integrate into the radiotherapy workflow to revolutionize cancer treatment. The overview of the radiotherapy workflow from patient setup and image segmentation to dose calculation and adaptive therapy, is described in Figure 6.

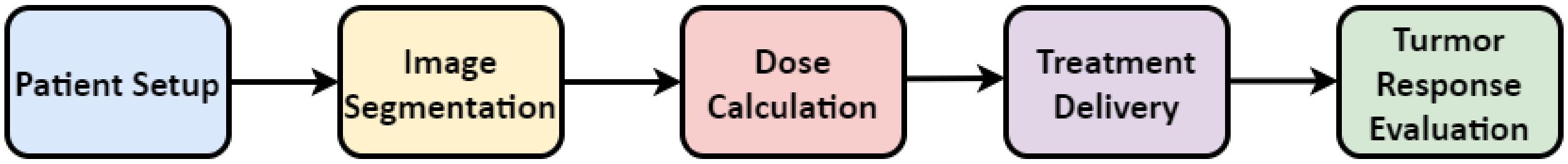

**Figure 6.** Overview of the radiotherapy workflow.

### 6.1 Patient Setup and Alignment

Accurate patient alignment is critical in radiotherapy to ensure that the radiation dose is precisely targeted at the tumor while sparing healthy tissues. Misalignment can lead to suboptimal treatment outcomes, increased radiation exposure to normal tissues, and potential harm to the patient. Precise alignment ensures consistent treatment delivery across multiple sessions, enhancing the overall effectiveness and safety of radiotherapy.

PCCT can enhance pre-treatment imaging for patient setup by providing high-resolution images with superior contrast. Although PCCTs are not currently integrated into treatment machines, their advanced detector technology allows for finer spatial resolution and better tissue differentiation in diagnostic settings. This high-resolution imaging aids in identifying anatomical landmarks more clearly during the simulation and treatment planning stages. For instance, small anatomical variations that could influence setup can be more accurately assessed during this pre-treatment phase, ensuring these are considered for positioning protocols on the treatment machine.

Recent developments, such as PCD-based multi-energy cone-beam CT (ME-CBCT), as explored by Hu et al. (Hu *et al* 2022b), show promise for improving on-board imaging in radiotherapy. ME-CBCT combines the advantages of photon-counting technology with cone-beam CT, offering improved contrast resolution and potentially more accurate patient alignment during daily treatments. The multi-energy capability allows for better tissue characterization and reduced artifacts, addressing a key limitation of conventional CBCT systems. Integrating PCCT-based CBCT into treatment machines could significantly enhance patient setup accuracy by providing higher-quality images during daily verification, aligning with the benefits seen in PCCT-based pre-treatment imaging.

Compared to conventional CT, PCCT offers advantages in pre-treatment imaging for radiotherapy planning. Conventional CT, though high in spatial resolution compared to modalities like MRI or PET, can sometimes struggle with specific small structures, such as the trabecular bone, particularly when contrast is low (Thomsen *et al* 2022). PCCT, with its higher resolution and improved CNR, allows for more precise identification of these structures, contributing to better pre-treatment planning and reducing the likelihood of setup errors when aligning the patient on the treatment machine. However, it’s important to note that on-board imaging systems, such as CBCT, are currently used for daily treatment verification and alignment, but future advancements like PCD-based ME-CBCT could further bridge this gap.

PCCT's superior image quality and contrast resolution suggest that it could produce more detailed Digitally Reconstructed Radiographs (DRRs) compared to conventional CT, potentially aiding in more accurate localization during treatment planning. However, further research is needed to confirm the full impact of PCCT-generated DRRs on daily alignment with on-board x-ray systems.

### 6.2 Image Segmentation (Target and OAR)

Accurate segmentation of targets and OAR is essential for effective radiotherapy planning. Segmentation in radiotherapy involves delineating the boundaries of the tumor (target) and nearby critical structures (OARs) to ensure that the maximum dose of radiation is delivered to the tumor while minimizing exposure to surrounding healthy tissues. Inaccurate segmentation can lead to suboptimal dose distribution, where parts of the tumor may receive insufficient radiation, or healthy tissues might be overexposed, leading to potential side effects and complications.

Moreover, precise segmentation is critical in advanced radiotherapy techniques such as intensity-modulated radiation therapy (IMRT) and stereotactic body radiation therapy (SBRT), where the radiation dose is shaped very closely to the target's contours. For example, studies involving liver have demonstrated that advanced imaging techniques such as DECT can significantly enhance segmentation accuracy (Chen *et al* 2020, Xu *et al* 2022). These techniques can then be adapted for use in radiotherapy, where precise tumor delineation is crucial to maximize therapeutic outcomes while minimizing exposure to healthy tissues in critical areas like the liver. Any errors in segmentation could lead to either an underdose in the tumor region, potentially compromising treatment effectiveness, or an overdose to surrounding healthy tissues, increasing the risk of radiation-induced toxicity.

PCCT can play a pivotal role in improving segmentation accuracy due to its ability to provide high-resolution, multi-energy images. Certain OARs, such as the brain and parotid glands, benefit from enhanced contrast and resolution, which aids in distinguishing sensitive structures (Shen *et al* 2023, Yu *et al* 2023). These images offer better tissue contrast and delineation, allowing for more precise identification and segmentation of tumors and OARs. PCCT's improved tissue differentiation capability, particularly for small or complex structures, can help overcome challenges that arise during radiotherapy planning in anatomically complex regions.

PCCT's multi-energy imaging capabilities enable the differentiation of tissues based on their specific energy-dependent attenuation properties. Aschenbrenner et al (Aschenbrenner *et al* 2017) demonstrated the feasibility of ultra-low-dose (ULD) photon-counting imaging for lung tumor position estimation, showing high accuracy with as few as five photons per pixel, potentially reducing radiation exposure while maintaining accurate tumor tracking. Marcus et al (Marcus *et al* 2018) showed PCCT's superiority over DECT in segmenting small renal stones, improving segmentation success (70% vs. 54.4%). Wang et al (Wang *et al* 2019) demonstrated that monoenergetic images at 80 keV improved OAR delineation in head-and-neck radiotherapy. Karino et al (Karino *et al* 2020) found that VMIs at 63 keV significantly improved brain metastasis segmentation during radiosurgery. Paakkari et al (Paakkari *et al* 2021) showed PCCT's capacity for energy-selective imaging, enhancing tissue segmentation with dual contrast agents in cartilage. Wang et al (Wang *et al* 2021b) developed a deep learning approach using DECT for automatic multi-organ segmentation in the head-and-neck, achieving a Dice similarity coefficient larger than 0.8 for complex structures. Baek et al (Baek *et al* 2024) demonstrated that using VMIs in PCCT significantly improved segmentation accuracy for the liver, pancreas, and spleen, particularly in low-dose scenarios.

These studies, summarized in Table 3, highlight the significant advantages of PCCT in improving segmentation accuracy across various clinical applications. By offering superior spatial resolution, enhanced contrast differentiation, and the ability to perform multi-energy imaging, PCCT allows for more precise identification and delineation of both tumors and OARs. This capability is particularly valuable in complex anatomical regions and in the context of advanced radiotherapy techniques, where accurate segmentation is crucial for optimizing treatment outcomes and minimizing the risk of radiation-induced side effects.

**Table 3.** Summary of Key Studies Demonstrating the Impact of Photon-Counting CT (PCCT) on Segmentation Accuracy Across Different Clinical Applications.

| **Study (Year)** | **Methodology Summary** | **Key Findings** | **Impact on Segmentation** |
|---|---|---|---|
| (Aschenbrenner *et al* 2017) | ULD photon-counting imaging for lung tumor position estimation. | Using just five photons per pixel could estimate tumor position within less than half a breathing phase. | Demonstrated PCCT's ability to reduce radiation dose while maintaining/improving tumor tracking accuracy, crucial for segmentation in radiotherapy. |
| (Marcus *et al* 2018) | Comparison of PCCT and DECT for renal stone characterization. | PCCT was superior in characterizing small renal stones (≤3 mm) compared to DECT, with a significant increase in successful segmentation (70.0% vs. 54.4%). | Highlighted PCCT's potential in enhancing segmentation accuracy for small structures, crucial for effective radiotherapy planning. |
| (Wang *et al* 2019) | Use of TwinBeam CT (TBCT) for monoenergetic imaging in head-and-neck OAR segmentation. | Identified that 80 keV provided the best CNR for brainstem, mandible, parotid glands, and spinal cord segmentation. | Highlighted TBCT's capability in enhancing OAR delineation through optimized monoenergetic imaging in head-and-neck radiotherapy. |
| (Karino *et al* 2020) | Analysis of optimal energy level of VMIs from contrast- | Found that a VMI at 63 keV provided the highest CNR, significantly | Emphasized the importance of selecting appropriate energy levels in spectral CT to enhance |
| | enhanced DECT for brain metastases. | improving lesion contrast and tumor delineation. | brain metastases segmentation. |
| (Paakkari *et al* 2021) | Simultaneous quantification of two contrast agents within articular cartilage using PCCT. | PCCT allowed for precise energy-selective imaging, enabling differentiation of tissue properties by quantifying cationic iodinated CA4+ and gadoteridol. | Demonstrated PCCT's capability in enhancing tissue segmentation, particularly in complex tissues like cartilage. |
| (Wang *et al* 2021b) | DECT-based deep learning model for multi-organ segmentation in the head-and-neck region. | DECT outperformed SECT in segmenting 19 organs, achieving Dice similarity coefficient (DSC) larger than 0.8 for major organs. | Demonstrated DECT's superiority in automatic segmentation, particularly for small, low-contrast structures in head-and-neck RT. |
| (Baek *et al* 2024) | UNet-based multi-organ segmentation in PCCT using VMIs. Evaluated the impact of noise reduction, material decomposition, and synthesized VMIs on segmentation accuracy in low- and high-dose cases. | Improved segmentation accuracy and training stability for liver, pancreas, and spleen; DSC increased from 0.933 to 0.95, with reduced standard deviation (0.066 to 0.047). | Demonstrated PCCT's ability to enhance segmentation accuracy using VMIs, particularly in low-dose scenarios. |

### 6.3 Radiation Dose Calculation

Dose calculation in radiotherapy is a critical process that involves calculating the optimal radiation dose distribution to target the tumor while minimizing exposure to surrounding healthy tissues. CT imaging is widely utilized in this process because it provides a detailed anatomical map of the patient's body, enabling precise visualization of both the tumor and surrounding structures. One of the key advantages of using CT in radiotherapy dose calculation is the straightforward conversion of Hounsfield Units (HU) into electron density. This electron density information is crucial for modeling how radiation interacts with tissues, ensuring that the prescribed dose is delivered effectively to the tumor while sparing healthy tissues.

One critical aspect of radiotherapy dose calculation is the accurate delineation of target volumes, which is often enhanced by the intravenous administration of iodine-based contrast agents during CT imaging. These agents improve the visibility of tumors, vessels nearby at-risk lymph nodes, and critical structures by increasing tissue attenuation, allowing for more precise targeting. However, the presence of iodine during imaging can lead to inaccuracies in dose calculations because the contrast agent is not present during the actual radiation delivery. This discrepancy can result in erroneous dose distributions, potentially compromising treatment effectiveness. To address this challenge, (Yamada *et al* 2014) proposed a novel framework that leverages dual-energy virtual unenhanced CT to improve dose calculation accuracy in the presence of iodine contrast agents with photons. Their study demonstrated that by using DSCT with enhanced spectral separation, it is possible to generate virtual unenhanced images that closely match the attenuation values of true unenhanced images. The researchers found that dose distributions calculated from these virtual unenhanced images were nearly equivalent to those from true unenhanced images, with pass rates exceeding 90%. Conversely, dose calculations based on contrast-enhanced images alone showed significant deviations, with pass rates of only 50–60%, highlighting the potential errors in dose estimation due to the high attenuation properties of iodine.

Yi-Qun et al (Yi-Qun *et al* 2016) investigated the use of DSCT with single-energy spectral imaging technology to further refine radiotherapy dose calculation for photons. Their study found that by processing images using single-energy spectrum imaging, it was possible to remove contrast agent artifacts, leading to more accurate dose calculations. For instance, when the iodine concentration was 30 g/100 mL, the deviation in dose measurement was significantly reduced from 5.95% in 80 kV images to 2.20% in spectral-fused images. This highlights the effectiveness of advanced spectral imaging techniques in enhancing the accuracy of radiotherapy planning, particularly when contrast agents are involved.

Another critical aspect of dose calculation is the accurate estimation of relative electron densities $\rho_e$ and Z_eff of tissues, which directly impact dose calculation accuracy, especially in proton therapy. Traditional methods involve CT scanning of tissue substitute materials (TSMs) to create HU– $\rho_e$ calibration curves, but these can introduce errors due to variations in tissue composition and x-ray beam spectra. In their study, (Tahmasebi Birgani *et al* 2018) proposed a more accurate approach by constructing an in-house phantom and applying dual-energy algorithms to scans at different kVp settings. Their methodology improved the differentiation of tissues with similar attenuation coefficients but different $\rho_e$ and Z_eff, leading to more precise characterization. The study demonstrated that using dual-energy algorithms combined with stoichiometric calibration reduced errors in $\rho_e$ calculation compared to traditional methods. For instance, the mean and standard deviation of the absolute difference in $\rho_e$ were significantly reduced when using 80–140 kVp and 100–140 kVp scans, compared to standard 120 kVp scans. Furthermore, a parametrization algorithm decreased the Z_eff discrepancy for tissues like the thyroid, achieving a residual error as low as 0.18 units. This low value represents the difference between the calculated Z_eff and the true value, providing context for the accuracy of the estimation.

McCollough et al (McCollough *et al* 2023) highlight, the dependence of CT numbers on x-ray beam spectra can limit the quantitative accuracy and standardization of these measurements, posing challenges in achieving robust and consistent applications in radiotherapy. PCCT technology addresses these limitations by offering inherent multi-energy capabilities, expanding material decomposition options, and improving spatial resolution and geometric quantification (Meloni *et al* 2023a). This makes PCCT a superior choice for dose calculation, as it allows for more precise and standardized computation.

These studies underscore the importance of incorporating advanced imaging technologies like PCCT into radiotherapy dose calculation, particularly for disease sites where precise differentiation is crucial. By improving the accuracy of dose calculations and reducing artifacts, these technologies ensure more precise targeting of the tumor involving complex anatomy or the use of intravenous contrast agents, ultimately leading to better patient outcomes.

**6.3.1 Proton Stopping Power Ratio**

One of the key advantages of PCCT in radiotherapy dose calculation is its ability to provide more accurate proton stopping power ratio (SPR) estimates. Accurate SPR values are essential for precise proton dose calculations. In conventional EICT, SPR is often estimated indirectly through conversion of HUs, which can introduce errors due to tissue inhomogeneities and the limitations of the conversion algorithm. PCCT, with its multi-energy imaging capability, allows for direct and more accurate measurement of SPR by leveraging its ability to differentiate between different tissue types based on their energy-dependent attenuation properties.

The improved accuracy of SPR estimates provided by PCCT directly impacts the precision of dose calculations in proton therapy. Accurate dose calculations ensure that the prescribed dose is delivered precisely to the tumor, maximizing therapeutic efficacy while minimizing exposure to healthy tissues. By reducing uncertainties in SPR estimates, PCCT contributes to more reliable and effective treatment plans, ultimately enhancing patient outcomes.

Over the years, various advanced techniques have been explored to improve SPR estimation, moving beyond the traditional SECT methods. These approaches include the use of DECT, MECT, and PCCT. Each method offers unique advantages, contributing to a more precise determination of SPR, which directly influences the accuracy of dose distribution in radiotherapy.

Studies such as Lalonde et al (Lalonde *et al* 2017) introduced Bayesian Eigen tissue decomposition for MECT, reducing SPR errors. Möhler et al (Möhler *et al* 2017) focused on DECT electron density, providing high accuracy for SPR estimations. Mei et al (Mei *et al* 2018) and Xie et al (Xie *et al* 2018) both validated dual-layer CT and stoichiometric DECT for improved SPR prediction. Taasti et al (Taasti *et al* 2018) and Saito (Saito 2019) further emphasized PCCT's superior accuracy in noise robustness and tissue characterization.

Simard et al (Simard *et al* 2019, 2020) adapted PCCT for Bayesian decomposition, achieving reduced RMS errors. Chang et al. (2022) introduced deep learning frameworks for SPR prediction with greater accuracy. Faller et al (Faller *et al* 2020) and Longarino et al (Longarino *et al* 2022a, 2022b) explored DLCT's clinical utility in particle therapy, reporting improvements in proton therapy precision by reducing beam range uncertainties.

Li et al (Li *et al* 2020) assessed MECT for stenosis quantification, demonstrating that it reduced partial volume and blooming effects, supporting more reliable SPR estimation in clinical practice. Nasmark and Andersson (Näsmark and Andersson 2021) introduced dual-energy VMIs, showing improved SPR accuracy in complex tissue environments, while Wang et al (Wang *et al* 2021a) developed a noise-robust learning method for accurate SPR predictions. Meng et al (Meng *et al* 2022) demonstrated the utility of DECT combined with metal artifact reduction (MAR) for SPR accuracy, particularly in challenging conditions involving metal artifacts.

Hu et al (Hu *et al* 2022a) validated PCCT for SPR using VMIs, achieving enhanced SPR accuracy for dose calculations. Zhu et al (Zhu *et al* 2023) focused on spectral CT for SPR prediction, showing dose deviations reduced by 2.57% in lung tumors. Longarino et al (Longarino *et al* 2023) also emphasized better SPR predictions with DLCT in patients with dental materials, improving clinical accuracy in particle therapy. Finally, Yang et al (Yang *et al* 2023) reviewed various DECT and MECT methods for SPR, highlighting their strengths for clinical proton therapy applications, while Zimmerman et al (Zimmerman *et al* 2023) and Han et al (Han *et al* 2024) explored algorithms and spectral separation for enhanced SPR in pediatric proton therapy cases. Fogazzi et al (Fogazzi *et al* 2024) compared DECT, PCCT, and proton CT (pCT), with PCCT showing superior image quality while pCT excelled in SPR accuracy. Larsson et al (Larsson *et al* 2024) demonstrated potential of combining PCCT with deep learning to reduce proton beam range uncertainty for enhancing the precision of dose calculation in radiotherapy. Zimmerman and Poludniowski (Zimmerman and Poludniowski 2024) demonstrated that PCCT provides narrower distributions, and improved SPR estimation accuracy compared to SECT and DECT in a head-and-neck phantom.

Table 4 summarizes key studies that have investigated various advanced CT methods, including DECT, MECT, and PCCT, for improving SPR estimation. These studies highlight the advancements in SPR accuracy achieved through different techniques and their potential clinical applications in proton therapy.

**Table 4.** Overview of Key Studies Comparing Proton Stopping Power Ratio (SPR) Estimation Accuracy.

| **Study (Year)** | **Methodology Summary** | **Key Findings** | **SPR Accuracy/Errors** |
|---|---|---|---|
| (Lalonde *et al* 2017) | Bayesian ETD for MECT data | Improved SPR estimation, reduced errors with more energy bins | RMS error 1.53% |
| (Möhler *et al* 2017) | DECT electron density estimation using alpha blending | High accuracy with minimal uncertainty, supports clinical application | Uncertainty 0.15% |
| (Mei *et al* 2018) | DLCT electron density estimation | High validity with errors within 1.79%, robust across different radiation doses | Error within 1.79% |
| (Xie *et al* 2018) | Stoichiometric DECT calibration for SPR | High accuracy, minimal deviation, suitable for clinical use | Mean 0.07%, SD 0.58% |
| (Taasti *et al* 2018) | PCCT for SPR estimation | Superior accuracy, robust against noise, promising for clinical application | RMSE 0.8%-1.0% |
| (Saito 2019) | ΔHU–ρe conversion using PCDs | Improved calibration accuracy, reduced errors, particularly in varying object sizes | Reduced calibration errors |
| (Simard *et al* 2019) | Bayesian decomposition with PCCT | Reduced RMS errors in SPR estimation compared to DECT | RMS error 1.4% (PCCT) |
| (Li *et al* 2020) | MECT for stenosis quantification | Improved accuracy in stenosis measurement, supports precise SPR estimation | Consistent and accurate measurements |
| (Simard *et al* 2020) | DECT vs. SPCCT in radiotherapy applications | SPCCT outperformed DECT in SPR accuracy, reduced RMSE | RMSE 0.89% (SPCCT) |
| (Faller *et al* 2020) | DLCT for SPR prediction in particle therapy | Reduced beam range uncertainty, improved precision in proton therapy | Mean accuracy 0.6% |
| (Näsmark and Andersson 2021) | Novel method using dual-energy virtual monoenergetic images | Improved SPR accuracy, robust against different scan parameters | SPR RMSE: 7.2% (lung), 0.4% (soft tissue), 0.8% (bone) |
| (Wang *et al* 2021a) | Noise-robust learning-based method to predict RSP maps from DECT | RSP prediction accuracy maintained in noisy environments, enhancing proton dose calculations | Mean square error 2.83% |
| (Hu *et al* 2022a) | PCCT for SPR estimation using VMIs | Significant improvement in SPR accuracy, particularly with VMIs at 60-180 keV | RSP accuracy: 1.27%-0.71% |
| (Longarino *et al* 2022a) | DLCT vs. SECT for SPR prediction in particle therapy | DLCT showed better agreement with measured values, improved dose accuracy | Mean deviation 0.7% (DLCT) vs. 1.6% (SECT) |
| (Longarino *et al* 2022b) | DLCT-based SPR prediction in heterogeneous anatomical regions | Clinically relevant range shifts, improved accuracy in complex regions | Range shifts: 0.4 mm to 2.1 mm |
| (Meng *et al* 2022) | MonoE CT images for SPR prediction in presence of metal artifacts | DECT combined with MAR algorithm provided robust SPR estimates | MAE of SPRw reduced to 1.05%-1.46% |
| (Chang *et al* 2022) | PIDL framework to derive accurate RSP maps from DECT images | PIDL method improved RSP accuracy for adult male, female, and child phantoms | Accuracy improvement by 3.3%-1.9% compared to ANN |
| (Zhu *et al* 2023) | Spectral CT to determine optimal energy pairs for SPR prediction | SPR estimation accuracy improved with optimal energy pairs in lung/brain tumors | Range difference: 1.84 mm in lung tumor, 2%/2 mm γ passing rate: 85.95% lung, 95.49% brain |
| (Longarino *et al* 2023) | Impact of dental materials on SPR prediction using DECT and DLCT | DLCT showed better SPR prediction accuracy with dental materials | Range deviation reduced to 0.2 mm |
| (Yang *et al* 2023) | Review of DECT and MECT methods for SPR estimation | Highlighted strengths and weaknesses, suggested improvements for clinical implementation | Emphasized reduction in SPR estimation uncertainties |
| (Zimmerman *et al* 2023) | Noise suppression algorithm for SPR estimation from DECT | Effective noise suppression, maintained spatial resolution, improved accuracy | Accuracy improvement in noisy environments |
| (Han *et al* 2024) | Impact of spectral separation on SPR prediction accuracy | Optimized spectral pairs significantly improved SPR estimation accuracy | Root-mean-squared error 0.12% |
| (Fogazzi *et al* 2024) | Comparison of DECT, PCCT, and pCT for SPR estimation | pCT was most accurate, but PCCT provided better image quality | MAPE: 0.28% (pCT), 0.80% (PCCT), 0.51% (DECT) |
| (Larsson *et al* 2024) | Deep learning-based SPR estimation from PCCT images using U-Net with simulated data (XCAT phantom) | Improved SPR prediction with PCCT data, reducing beam range uncertainties in proton therapy | RMSE 0.26%-0.41% (PCCT) vs. 0.40%-1.30% (SECT) and 0.41%-3.00% (DECT) |
| (Zimmerman and Poludniowski 2024) | Comparison of PCCT, DECT, and SECT for material characterization and SPR estimation in a head-and-neck phantom | PCCT outperformed DECT and SECT in most metrics, with narrower distributions and lower RMSDs for RED, EAN, and SPR across most materials | RMSD for SPR: PCCT lowest for most materials; SECT performed poorest, especially in complex scenarios |

### 6.4 Metal Artifact Reduction (MAR)

Metal artifacts present a significant challenge in diagnostic imaging, as they can obscure critical anatomical structures and distort surrounding tissue visualization. These artifacts are typically caused by high-density materials such as dental fillings, orthopedic implants, and surgical clips, which interfere with x-ray beams and result in streaks or dark shadows on conventional EICT images. This degradation in image quality can hinder accurate diagnostic interpretation and pose challenges for various clinical applications, including radiotherapy, surgical planning, and diagnostic evaluations.

PCCT offers advanced techniques for MAR, leveraging its multi-energy imaging capabilities to differentiate between high-density metals and surrounding tissues. By counting individual photons and sorting them into specific energy bins, PCCT can mitigate the effects of metal artifacts more effectively than conventional EICT, thereby reducing metal artifacts. This technology is particularly valuable in cases where metal implants are present, such as hip replacements, dental work, or prosthetics, as it reduces blooming artifacts and distortion caused by high-density materials. As a result, PCCT enables clinicians to visualize and analyze tissues and structures more accurately, even in challenging environments with metallic elements (Selles *et al* 2024, O'Connell *et al* 2024).

Several studies have highlighted the benefits of PCCT in reducing metal artifacts, summarized in Table 5. For example, Brook et al (Brook *et al* 2012) evaluated the use of spectral CT with MAR software for reducing artifacts associated with gold fiducial seeds, finding that MAR reconstruction significantly improved image quality near the seeds. This improvement is particularly relevant for patients with both gold markers and I-125 seeds in LDR prostate brachytherapy, where it can be difficult to distinguish larger gold markers from clustered I-125 seeds. PCCT may help address this challenge by providing clearer differentiation between different materials.

Li et al (Li *et al* 2020) demonstrated that MECT reduces partial volume and blooming effects, allowing for more accurate measurements, even in areas with calcifications or other dense materials. Allphin et al (Allphin *et al* 2023) explored the benefits of combining different PCDs in a micro-CT system, showing improved spectral performance and significantly reduced artifacts. Stein et al (Stein *et al* 2023) systematically evaluated the impact of PCCT on imaging small vessel stents, confirming superior image quality and reduced artifacts compared to conventional EICTs. The advancements in MAR techniques provided by PCCT may offer significant clinical benefits, not only in radiotherapy but also across other fields such as diagnostic radiology, interventional procedures, and post-surgical follow-up. With its ability to mitigate the impact of metal artifacts, PCCT may enhance the accuracy and reliability of imaging, improving patient outcomes and supporting more effective clinical decision-making. Even in non-patient scenarios, like QA equipment and phantom imaging, metal artifacts, such as distorted ball bearings and streaks in detector arrays, present challenges. PCCT's potential for reducing metal artifacts may provide notable improvements in quality assurance processes.

**Table 5.** Overview of Studies on Metal Artifact Reduction with Photon-Counting CT.

| **Study (Year)** | **Methodology Summary** | **Key Findings** | **Impact on MAR** |
|---|---|---|---|
| (Brook *et al* 2012) | Evaluation of spectral CT with MAR software for reducing artifacts from gold fiducial seeds. | MAR-reconstructed images showed improved tumor visibility near fiducial seeds, with reduced blooming artifacts. | Demonstrated the effectiveness of MAR in enhancing image quality and improving visualization near high-density materials. |
| (Li *et al* 2020) | MECT for stenosis quantification, focusing on reducing partial volume and blooming effects. | MECT images provided accurate and reproducible measurements, with significantly reduced errors compared to SECT, especially in calcified areas. | Highlighted MECT's ability to reduce artifacts and improve measurement accuracy in complex imaging scenarios. |
| (Allphin *et al* 2023) | Performance evaluation of two PCDs for micro-CT imaging of phantoms. | Combining GaAs and CdTe detectors improved spectral performance and significantly reduced artifacts compared to single detector systems. | Showed that multiple PCDs enhance MAR, leading to clearer imaging and better diagnostic accuracy. |
| (Stein *et al* 2023) | Systematic investigation of PCCT's impact on small vessel stent assessment. | PCCT provided superior image quality with reduced artifacts compared to EID-CT, particularly in small vessel stent imaging. | Confirmed that PCCT's advanced imaging capabilities significantly improve MAR, essential for accurate imaging of stents and implants. |

## 6.5 Tumor Response Evaluation/Outcome Modeling

Monitoring tumor response during and after radiotherapy is a vital aspect of evaluating treatment efficacy and optimizing patient care. Accurate assessment of how a tumor responds to radiation not only provides insights into the effectiveness of the therapy but also guides potential modifications to the treatment plan. This ongoing evaluation is crucial for ensuring that the tumor is being targeted effectively while minimizing damage to surrounding healthy tissues.

PCCT offers significant advancements in tumor response evaluation due to its ability to produce high-resolution, multi-energy images. These images provide detailed insights into tumor morphology, allowing clinicians to track changes in tumor size, shape, and density over time. The multi-energy capabilities of PCCT may also enable the differentiation of various tissue types based on their specific attenuation properties, making it easier to distinguish between viable tumor tissue and areas of necrosis or fibrosis that may occur as a result of treatment.

Moreover, PCCT's ability to generate quantitative imaging data plays a crucial role in outcome modeling in radiotherapy. By analyzing the detailed imaging features provided by PCCT, clinicians can develop predictive models that correlate specific imaging biomarkers with treatment outcomes. For example, changes in tumor volume, density, and contrast enhancement patterns observed through PCCT can be used to predict the likelihood of treatment success or the potential for tumor recurrence.

This quantitative approach enables personalized treatment planning, where adjustments to the radiation dose, treatment duration, or even the introduction of additional therapies can be made based on the observed tumor response. Such data-driven decisions can significantly improve patient outcomes by ensuring that the treatment remains effective throughout its course and by allowing for timely interventions when necessary.

Furthermore, the ability of PCCT to provide longitudinal imaging data with minimal artifacts and high contrast resolution enhances the precision of tumor response evaluation. This is particularly important in ART, where real-time adjustments to the treatment plan are made based on changes in tumor morphology. The use of PCCT in this context not only enhances the accuracy of tumor response assessment but also supports the implementation of more dynamic and responsive treatment strategies.

Several studies have demonstrated the utility of advanced CT technologies like spectral CT and PCCT in evaluating tumor response. Hu et al (Hu *et al* 2014) explored spectral CT's ability to monitor treatment response in pancreatic carcinoma xenografts, correlating reduced iodine concentration with treatment efficacy. Aoki et al (Aoki *et al* 2016) found that lower iodine density in lung tumors, as evaluated by DECT, predicted worse local control post-SBRT. Similarly, Al-Najami et al (Al-Najami *et al* 2017) and Lapointe et al (Lapointe *et al* 2017) demonstrated DECT's utility in quantifying tumor regression and lung function preservation in rectal and lung cancers, respectively.

Further extending the potential of PCCT, Nicol et al (Nicol *et al* 2019) and Fehrenbach et al (Fehrenbach *et al* 2019) highlighted its applications in cardiovascular imaging and non-small cell lung cancer (NSCLC), suggesting these techniques could be applied similarly in oncology for more accurate response evaluations. Liao et al (Liao *et al* 2022) and Inoue et al (Inoue *et al* 2022) explored spectral CT's role in predicting treatment responses in nasopharyngeal carcinoma and lung disease, while Salas-Ramirez et al (Salas-Ramirez *et al* 2022) and Fu et al (Fu *et al* 2023) showcased DECT's use in bone marrow quantification and osteosarcoma imaging.

In other areas, Tong et al (Tong *et al* 2024) developed a nomogram combining spectral CT data with clinical characteristics to predict lymphovascular invasion in gastric cancer, while Li et al (Li *et al* 2024b) assessed DECT parameters for predicting radiotherapy sensitivity in nasopharyngeal carcinoma. Finally, Bellin et al (Bellin *et al* 2024) reviewed DECT's potential in improving diagnostic accuracy for renal cell carcinoma, suggesting that similar benefits may extend to tumor response assessments in radiotherapy. These advancements could translate to similar benefits in evaluating tumor response in radiotherapy, showcasing the broad applicability of these imaging technologies. Surov et al (Surov *et al* 2024) demonstrated that normalized iodine concentration (NIC) values derived from PCCT could predict treatment response in rectal cancer, with high sensitivity and specificity, emphasizing its potential for guiding therapy decisions.

Table 6 summarizes key studies that investigate the use of spectral CT and PCCT for tumor response evaluation across various cancer types. These studies demonstrate the ability of advanced CT technologies to monitor vascular changes, predict tumor progression, and assess treatment outcomes more accurately, supporting personalized treatment strategies in radiotherapy and oncology.

**Table 6.** Summary of Studies on Tumor Response Evaluation Using Spectral and Photon-Counting CT.

| Study (Year) | Methodology Summary | Key Findings | Impact on Tumor Response Evaluation |
|---|---|---|---|
| (Hu *et al* 2014) | Spectral CT for monitoring therapeutic response to 125I brachytherapy in pancreatic carcinoma xenografts. | nIC correlated with microvessel density, indicating spectral CT's potential in non-invasive treatment response evaluation. | Demonstrated spectral CT's utility in assessing vascular changes post-therapy. |
| (Aoki *et al* 2016) | DECT for evaluating iodine density in lung tumors treated with SBRT. | Lower iodine density was associated with worse prognosis, reflecting hypoxic cell populations. | Highlighted DECT's potential in predicting tumor radioresistance. |
| (Lapointe *et al* 2017) | DECT for retrieving lung function information for radiotherapy planning, compared with SECT and SPECT/CT. | Strong correlation between DECT and SPECT/CT functional data, showing DECT's utility in lung function preservation. | Provided insights into lung function that are crucial for functional tissue sparing during RT. |
| (Al-Najami *et al* 2017) | DECT for quantifying tumor regression post-neoadjuvant therapy in rectal cancer. | DECT parameters correlated with pathological regression, suggesting non-invasive quantification of tumor response. | Potential use of DECT for guiding treatment adjustments based on tumor regression. |
| (Nicol *et al* 2019) | Review of photon-counting CT's future role in cardiovascular imaging. | Discussed advancements in photon-counting CT and AI-driven analyses for more accurate diagnostic outcomes. | Paralleled the use of photon-counting CT in oncology for enhancing tumor response evaluation. |
| (Fehrenbach *et al* 2019) | Spectral CT for assessing NSCLC response to chemoradiotherapy. | Higher iodine content was associated with tumor progression, suggesting IC as a predictive biomarker. | Suggested that spectral CT-derived biomarkers could predict treatment response. |
| (Liao *et al* 2022) | Spectral CT-based nomogram for predicting ICT response in NPC. | The nomogram showed high predictive accuracy and could guide personalized treatment strategies. | Demonstrated the potential of spectral CT in personalized treatment planning. |
| (Inoue *et al* 2022) | Comparison of PCD-CT and EID-CT for diagnosing UIP. | PCD-CT provided better image quality and improved reader confidence, suggesting potential applications in tumor response evaluation. | Indicated that PCD-CT could improve confidence in evaluating tumor response. |
| (Salas-Ramirez *et al* 2022) | DECT for quantifying bone marrow components in dosimetry. | DECT accurately quantified marrow components, which could be crucial for evaluating bone involvement in tumors. | Provided a method for evaluating bone marrow involvement in tumor response. |
| (Fu *et al* 2023) | DECT with BiOI nanosheets as a contrast agent for osteosarcoma imaging. | Improved tumor visualization and guided radiotherapy, highlighting the potential of DECT in enhancing tumor response evaluation. | Enhanced DECT's specificity for tumor response evaluation. |
| (Tong *et al* 2024) | Nomogram based on spectral CT and clinical data to predict LVI in gastric cancer. | High predictive accuracy of the nomogram emphasized spectral CT's role in assessing tumor response. | Supported the use of spectral CT for personalized tumor response assessment. |
| (Li *et al* 2024b) | DECT parameters for predicting radiotherapy sensitivity in NPC. | DECT parameters were effective predictors, aiding in treatment planning adjustments. | Reinforced the predictive power of DECT in radiotherapy sensitivity. |
| (Bellin *et al* 2024) | Review of recent advances in renal cell carcinoma imaging. | Discussed DECT's potential in improving diagnostic accuracy and reducing additional imaging needs. | Suggested similar benefits in evaluating tumor response in radiotherapy. |
| (Surov *et al* 2024) | Pilot study correlating iodine concentration (IC) from PCCT with histopathology and treatment response in rectal cancer (RC) | Higher normalized IC (NIC) values were associated with lymphovascular invasion | NIC values demonstrated high inter-reader agreement (ICC=0.93) and predictive accuracy for treatment response (AUC=0.85) |

### 6.6 Adaptive Radiotherapy (ART)

Adaptive radiotherapy (ART) involves adjusting the treatment plan in real-time or offline between fractions based on changes in patient anatomy or tumor size during treatment. PCCT's high-resolution, multi-energy imaging capabilities make it ideal for both approaches by providing detailed images that help monitor anatomical changes, ensuring precise treatment adjustments. This real-time adaptability allows the radiation dose to remain optimally targeted to the tumor, enhancing both treatment efficacy and patient safety.

Deformable registration plays a key role in ART, as it aligns images taken at different times to track anatomical changes. PCCT's superior imaging quality enhances the accuracy of deformable registration, ensuring that shifts in patient anatomy are precisely tracked. Accurate registration is critical for delivering the correct dose to the tumor while minimizing radiation exposure to surrounding healthy tissues, ultimately improving ART's effectiveness.

Hu et al (Hu et al 2022b) developed a PCD-based ME-CBCT system on a preclinical small animal irradiator to enhance the accuracy of material differentiation and dose calculation, comparing its performance to conventional flat-panel detector (FPD)-based CBCT. The study involved mounting a PCD onto an existing irradiator platform and utilizing a 100 kVp x-ray beam with three optimized energy thresholds for SNRs. The results demonstrated that using PCD-based ME-CBCT significantly improved dose calculation accuracy, reducing the mean relative error in bone regions from 49.5% to 16.4% and in soft tissue regions from 7.5% to 6.9%. While this study focused on a preclinical irradiator platform, PCCT could offer similar benefits in offline ART, where treatment adaptations occur between fractions. Additionally, PCCT's advanced imaging may also be applicable to lower-energy procedures, such as electronic brachytherapy (50 kVp), which is used in spine intraoperative RT. This would allow for better treatment precision in such specialized clinical applications, further expanding PCCT's utility in adaptive radiotherapy scenarios.

## 7. Radiomics and Deep Learning Using PCCT

Radiomics and deep learning are two rapidly evolving fields in medical imaging, offering powerful tools for improving diagnosis, treatment planning, and patient outcomes. Radiomics involves extracting large numbers of quantitative features from medical images (Gillies *et al* 2016), which can then be used for predictive modeling and personalized treatment planning. Deep learning, on the other hand, leverages artificial intelligence to automate and enhance the interpretation of complex imaging data (LeCun *et al* 2015). PCCT serves as an ideal platform for both technologies due to its ability to capture high-resolution, multi-energy images, providing a wealth of data for feature extraction and deep learning analysis.

### 7.1 Radiomics in PCCT

Radiomics involves the extraction of a vast array of quantitative features from medical images, which can be used to build predictive models that assist in patient outcome predictions. Yang et al (Yang *et al* 2018) demonstrated that DECT-derived features like iodine concentration could predict therapeutic outcomes in laryngeal cancers. Allphin et al (Allphin *et al* 2022) found that PCCT-derived radiomics features offered higher accuracy for tumor stratification than EID. Ayx et al (Ayx *et al* 2022) compared PCCT and EICT in myocardial imaging, observing significant differences in higher-order texture features due to PCCT's enhanced resolution. Ter Maat et al (Ter Maat *et al* 2023) suggested that PCCT’s spectral data could enhance radiomics models in personalized treatment strategies.

### 7.2 Deep Learning in PCCT

PCCT's multi-energy imaging capabilities provide an ideal dataset for training deep learning algorithms, offering improved accuracy in image segmentation, diagnosis, and treatment planning. Wen et al (Wen *et al* 2022) demonstrated DL's potential in differentiating central lung cancer from atelectasis, significantly improving tumor delineation. Similarly, Jacobsen et al (Jacobsen *et al* 2020) and Van Der Werf et al (Van Der Werf *et al* 2022) highlighted the quantitative benefits of PCCT for various applications, stressing the need for advanced DL models. Wang et al (Wang *et al* 2023) introduced DL models with PCCT for predicting temperature changes during ablation, enhancing procedural accuracy. Ge et al (Ge *et al* 2024) and Sun et al (Sun *et al* 2024) further integrated PCCT-derived features into DL models, improving predictive accuracy for conditions such as LVI in gastric cancer and colorectal adenocarcinoma staging. Sidky and Pan (Sidky and Pan 2024) showcased DL's superiority in spectral CT image reconstruction, while Alves et al (Alves *et al* 2024) emphasized PCCT's advantages over EICT for feature extraction and deep learning applications.

Table 7 summarizes key studies that explore the intersection of PCCT with radiomics and deep learning, highlighting the various clinical applications and the potential benefits these technologies offer in improving tumor assessment and treatment outcomes.

The integration of radiomics and deep learning with PCCT holds great promise for advancing the field of radiotherapy. By leveraging PCCT's spectral imaging capabilities, these technologies enable more accurate diagnosis, improved treatment planning, and better patient outcomes. As studies continue to demonstrate the advantages of PCCT over conventional CT, particularly in personalized medicine, the future of radiomics and deep learning in radiotherapy looks bright.

**Table 7.** Key Studies Investigating the Integration of Spectral and Photon-Counting CT with Radiomics and Deep Learning.

| Study | Methodology Summary | Key Findings | Relevance to PCCT |
|---|---|---|---|
| (Yang *et al* 2018) | Dual-energy CT to predict therapeutic effects in advanced LHSCC. | DECT-derived parameters, such as λ(HU), helped identify patients with CR vs NCR. | Demonstrates the potential of DECT in radiomics for cancer treatment assessment. |
| (Allphin *et al* 2022) | Spectral micro-CT with nanoparticle contrast for tumor stratification. | PCD-based radiomic features showed higher accuracy in differentiating tumors based on lymphocyte burden. | PCCT-derived radiomics can enhance tumor stratification for personalized cancer therapy. |
| (Ayx *et al* 2022) | Comparison of radiomics features between PCCT and EICT in myocardial imaging. | Higher-order texture features showed significant differences between PCCT and EICT, while first-order features were comparable. | Demonstrates PCCT's superior spatial resolution and impact on radiomics feature extraction, supporting its use in quantitative imaging. |
| (Ter Maat *et al* 2023) | Radiomics to predict outcomes of checkpoint inhibitor treatments in melanoma. | Radiomics had moderate predictive value but did not improve on clinical models. | Highlights the need for advanced radiomics and deep learning integration with PCCT. |
| (Wen *et al* 2022) | Double-layer spectral CT to differentiate lung cancer from atelectasis. | Spectral images enhanced tumor border delineation, improving staging accuracy. | Demonstrates improved tumor identification and segmentation using spectral CT and deep learning. |
| (Jacobsen *et al* 2020) | Review of MECT applications for quantitative imaging. | PCCT offers improved quantitative metrics, including bone mineral density and disease biomarkers. | PCCT is poised to enhance quantitative imaging in radiotherapy and diagnostics. |
| (Van Der Werf *et al* 2022) | PCCT for improved CAC scoring compared to conventional CT. | PCCT provided better CAC detection and quantification due to improved spatial resolution. | PCCT is superior for capturing fine anatomical details, essential for cardiovascular assessments. |
| (Wang *et al* 2023) | Deep learning and PCCT for real-time 3D temperature visualization during ablation. | PCCT enabled accurate thermometry for tumor ablation procedures. | Highlights the utility of PCCT in procedural guidance and ablation therapy. |
| (Ge *et al* 2024) | Deep learning model combining clinical markers and spectral CT for LVI/PNI prediction in gastric cancer. | Spectral CT parameters and clinical data integration improved preoperative LVI/PNI prediction. | Demonstrates the benefit of integrating spectral CT data with deep learning for cancer prognosis. |
| (Sun *et al* 2024) | DLCT with deep learning for colorectal adenocarcinoma staging. | ECV from DLCT improved pT staging with high diagnostic accuracy. | Suggests PCCT's potential in accurate non-invasive cancer staging. |
| (Sidky and Pan 2024) | AAPM Grand Challenge for deep learning-based spectral CT reconstruction. | Deep learning models achieved highly accurate reconstructions, surpassing traditional methods. | PCCT with deep learning offers superior image quality and reconstruction accuracy. |
| (Alves *et al* 2024) | Comparison of PCCT and EICT for feature extraction in radiomics and deep learning. | PCCT's superior spatial resolution and enhanced material decomposition make it a more robust platform for feature extraction. | Confirms PCCT's advantages over EICT in feature extraction for radiomics and deep learning applications. |

## 8. Internal Radiotherapy: Brachytherapy and Radiopharmaceutical Therapy (RPT)

Internal radiotherapy, including brachytherapy and radiopharmaceutical therapy (RPT), relies on the precise delivery of radiation to tumors while minimizing damage to surrounding healthy tissues. PCCT has the potential to revolutionize these therapies by providing high-resolution, multi-energy images that enhance both dosimetry calculations and real-time treatment monitoring.

In brachytherapy, both LDR and high-dose-rate (HDR) methods involve implanting radioactive sources directly into or near the tumor for targeted radiation delivery. However, metal artifacts from both the radioactive seeds in LDR and metal applicators used in HDR brachytherapy can obscure critical anatomical details, complicating dosimetry and treatment planning. PCCT could address these challenges by providing high-resolution, artifact-reduced images, improving visualization and accuracy for precise dose delivery in internal radiotherapy. Current research highlights the potential of PCCT to address these challenges by providing clearer, artifact-free images. Hu et al (Hu *et al* 2014) demonstrated that spectral CT could effectively track therapeutic response to $^{125}$I interstitial brachytherapy in a pancreatic carcinoma model showing a significant correlation between iodine concentration and microvessel density in treated tumors. Additionally, Yang et al (Yang *et al* 2015) investigated the use of spectral CT combined with MAR software (MARS) to optimize imaging quality around $^{125}$I seeds used in liver brachytherapy. Their findings indicated that using monochromatic images at 75 keV, with the aid of MARS, substantially reduced artifacts, allowing for better tumor visibility and accurate treatment evaluation supporting the idea that PCCT could further enhance image quality in brachytherapy by offering even higher resolution and better metal artifact reduction.

RPT, which involves the administration of radioactive substances to target tumors, also benefits from the capabilities of PCCT. Accurate tracking and quantification of radiopharmaceutical uptake in both the tumor and surrounding tissues are essential for effective treatment planning and monitoring. The work of Liu et al (Liu *et al* 2016) on gemstone spectral imaging (GSI) and metal artifact reduction demonstrated that GSI could effectively reduce artifacts from $^{125}$I seeds, improving the CNR and visibility of adjacent tissues, indicates that PCCT's multi-energy imaging capabilities could further improve the precision of dosimetry calculations and enhance the effectiveness of RPT.

The future of internal radiotherapy lies in personalized treatment approaches, and PCCT holds the key to enabling such precision. By providing clearer, artifact-free images and allowing for more accurate tracking of radioactive seeds and radiopharmaceuticals, PCCT can enhance dosimetry and treatment monitoring. This would lead to highly individualized treatment plans that adjust dynamically to the patient's specific anatomy and response to therapy, ultimately improving treatment outcomes and minimizing complications.

Table 8 summarizes key studies that have investigated the impact of spectral CT and related technologies on improving image quality and artifact reduction in internal radiotherapy, demonstrating the potential application of PCCT in this domain.

**Table 8.** Key Studies on Artifact Reduction in Internal Radiotherapy Using Spectral CT and Potential Applications of PCCT.

| Study | Methodology Summary | Key Findings | Relevance to PCCT |
|---|---|---|---|
| (Hu *et al* 2014) | Spectral CT to evaluate therapeutic response to $^{125}$I brachytherapy in pancreatic carcinoma. | Significant correlation between iodine concentration and microvessel density, showing potential for evaluating treatment response. | Demonstrates potential of spectral CT in improving dosimetry and treatment monitoring in brachytherapy. |
| (Yang *et al* 2015) | Investigated optimal energy settings for artifact reduction from $^{125}$I seeds in liver brachytherapy using spectral CT with MARS. | 75 keV with MARS significantly reduced artifacts, improving diagnostic image quality. | PCCT could offer superior artifact reduction for enhanced visibility in brachytherapy treatments. |
| (Liu *et al* 2016) | GSI CT with MARs for artifact reduction in patients with $^{125}$I seed implantation. | 70 keV images provided best tissue contrast and reduced metal artifacts, improving image quality around the seeds. | PCCT has the potential to further enhance metal artifact reduction in internal RT. |

## 9. CT Dose Reduction

One of the most notable advantages of PCCT over conventional EICT is its potential for significant radiation dose reduction, which is particularly relevant in RT settings. In treatments such as proton therapy, frequent scans are necessary to verify patient setup or adapt to anatomical changes. PCCT's lower-dose imaging capability, while maintaining or enhancing image quality, supports these needs without adding significant radiation burden. This aligns with the growing efforts in clinical care to reduce exposure, particularly as CT scan utilization has dramatically increased, where approximately 80 million scans are performed annually in the United States, (Patel *et al* 2017). Despite advances in dose-reduction techniques, such as iterative reconstruction, automatic exposure control, and electrocardiography-triggered imaging, radiation exposure remains a significant concern. These techniques, while effective, have not fully addressed the cumulative risk associated with repeated imaging, particularly in vulnerable populations such as pediatric and oncology patients.

PCCT addresses these concerns by minimizing electronic noise and enabling x-ray photon energy weighting. This allows for reduced image noise at the same x-ray exposure compared to conventional CT scanners, leading to an overall reduction in radiation dose. Krauss et al (Krauss *et al* 2015) highlighted tin filtration's ability to improve dose efficiency in DECT, while Zeng et al (Zeng *et al* 2016) demonstrated advanced denoising filters reducing noise and allowing dose reductions. Gao et al (Gao *et al* 2016) showed an 11% radiation dose reduction in CT angiography with PCCT, while Symons et al (Symons *et al* 2017a, 2017b, 2017c) presented up to 10% dose savings in lung cancer screening and low-dose chest CT, with reduced noise and improved image quality.

Apfaltrer et al (Apfaltrer *et al* 2020) achieved a 25% dose reduction in urolithiasis imaging, with Naveed et al (Naveed *et al* 2021) showing PCCT could differentiate blood from contrast in embolization, reducing the need for follow-up imaging. Graafen et al (Graafen *et al* 2022) and Woeltjen et al (Woeltjen *et al* 2022) confirmed significant dose reductions in lung imaging, with Graafen showing a 50% reduction. Jungblut et al (Jungblut *et al* 2022) reported a 66% dose reduction for systemic sclerosis imaging, while Donuru et al (Donuru *et al* 2023) saw a nearly 40% lower dose in non-contrast chest CT. Dirrichs et al (Dirrichs *et al* 2023) highlighted superior pediatric cardiovascular imaging at comparable doses, and Dettmer et al (Dettmer *et al* 2024) achieved a chest CT dose similar to a chest x-ray with PCCT.

Milos et al (Milos *et al* 2024) reported a tenfold reduction in radiation dose for lung transplant patients, maintaining diagnostic accuracy for lung abnormalities, while Huflage et al (Huflage *et al* 2024) showed ultra-high-resolution PCCT maintained superior image quality at no dose disadvantage, benefiting repeated scans for high-risk patients.

Table 9 summarizes several studies comparing the radiation dose and image quality between PCCT and conventional EICT. Taken together, these findings consistently highlight PCCT's ability to reduce radiation doses by 30% to 60%, depending on the clinical application, while maintaining or even enhancing image quality. This makes PCCT a highly promising tool for improving patient safety across a wide range of imaging tasks, particularly in fields like oncology, pediatrics, and long-term follow-up scenarios where dose reduction is paramount.

**Table 9.** Comparison of Radiation Dose and Image Quality between PCCT and EICT.

| Study | Study Objective | PCCT vs. EICT (Radiation Dose) | Key Findings |
|---|---|---|---|
| (Krauss *et al* 2015) | Evaluate DECT with different voltage combinations and tin filtration | N/A | Tin filtration (80/150 Sn) improved noise reduction and dose efficiency, with better spectral separation for DE CT imaging. |
| (Zeng *et al* 2016) | Apply aviNLM filter for SCT image restoration and dose reduction | N/A | aviNLM filter suppressed noise and artifacts, enhancing image quality for SCT while allowing for potential dose reduction. |
| (Gao *et al* 2016) | Assess DE spectral CT for low-iodine intake and dose reduction in CTA | 9.09 mSv (PCCT) vs. 10.17 mSv (EICT) | 11% reduction in radiation dose and 22.86% reduction in iodine intake with PCCT while maintaining diagnostic quality. |
| (Symons *et al* 2017a) | Compare PCCT with EID-CT for low-dose lung cancer screening | Up to 10% lower (PCCT) | PCCT provided better HU stability and lower noise (up to 10% reduction), improving image quality at reduced dose. |
| (Symons *et al* 2017b) | Evaluate in vivo simultaneous material decomposition of multiple contrast agents | N/A | PCCT allowed simultaneous material decomposition, potentially reducing the need for multiphase CT and lowering radiation exposure. |
| (Symons *et al* 2017c) | Investigate low-dose PCCT vs. EICT in chest CT | N/A | PCCT achieved superior image quality and lower image noise, particularly for lung nodule detection, across different BMI groups. |
| (Apfaltrer *et al* 2020) | Assess dose reduction in urolithiasis using a low-dose stone-targeted DECT protocol | 3.34 mSv (PCCT) vs. 4.45 mSv (EID) | PCCT allowed for 24.9% dose reduction while maintaining high diagnostic image quality in stone-targeted dual-energy CT scans. |
| (Naveed *et al* 2021) | Use DECT to differentiate contrast from blood after MMA embolization | N/A | DECT, including PCCT, helped differentiate between contrast and blood, reducing the need for additional imaging and reducing radiation exposure. |
| (Graafen *et al* 2022) | Compare intrapatient radiation dose and image quality of PCD-CT vs. EID-CT for lung HRCT | 0.9 mGy (PCCT) vs. 1.8 mGy (EID) | PCCT achieved a 50% dose reduction while maintaining or improving image quality, demonstrating superior SNR and subjective image quality. |
| (Woeltjen *et al* 2022) | Evaluate PCCT for low-dose high-resolution lung CT | N/A | PCCT images exhibited better quality and lower noise, while reducing radiation dose compared to EID-CT. |
| (Jungblut *et al* 2022) | Examine dose reduction potential of PCCT in systemic sclerosis and ILD | 66% lower dose with PCCT | PCCT achieved a 66% dose reduction compared to EID-CT, without compromising image quality or diagnostic accuracy for ILD. |
| (Donuru *et al* 2023) | Compare non-contrast chest CT from PCCT and EID-CT | 4.710 mGy (PCCT) vs. 7.80 mGy (EID) | PCCT reduced radiation dose by nearly 40%, while 3 out of 5 radiologists preferred PCCT images for better image quality. |
| (Dirrichs *et al* 2023) | Assess radiation dose and image quality of PCCT vs. | 0.50 mSv (PCCT) vs. 0.52 mSv (DSCT) | PCCT provided superior image quality (higher SNR and CNR) with the same dose level as |
| | DSCT in pediatric cardiovascular imaging | | DSCT, improving diagnostic performance in pediatric cardiovascular imaging. |
| (Dettmer *et al* 2024) | Establish a ULD chest CT protocol using PCCT, matching CXR dose levels | 0.11 mSv (ULD-CT using PCCT) | PCCT matched CXR dose levels while providing significant diagnostic changes in 41% of cases, with a low-dose chest CT protocol. |
| (Milos *et al* 2024) | Evaluate ultralow-dose PCCT for lung transplant follow-up | 0.26 mSv (ULD1) vs. 1.41 mSv (LD) | PCCT allowed for a 10-fold radiation dose reduction while maintaining over 70% detection accuracy for lung abnormalities. |
| (Huflage *et al* 2024) | Investigate dose burden of PCCT in lung CT with UHR and OBTCM | 0.34-3.99 mSv (PCCT in UHR mode) | UHR mode provided superior image sharpness without a dose disadvantage compared to standard mode. OBTCM offered moderate dose savings. |

## 10. Discussion

PCCT represents a significant breakthrough in imaging technology. By providing enhanced spatial and contrast resolution, multi-energy imaging capabilities, and a reduced radiation dose, PCCT holds promise for transforming imaging applications, particularly in the field of radiotherapy. This discussion explores the clinical applications of PCCT, its benefits for a diverse range of patient populations, and forthcoming innovations. It also addresses challenges such as cost considerations and the necessity for radiotherapy-specific clinical trials to fully realize PCCT's potential.

### 10.1 PCCT Advances in Cancer Radiotherapy

In cancer radiotherapy, PCCT can enhance tumor targeting and treatment planning significantly. By delivering high-resolution, multi-energy images, PCCT sharpens the segmentation accuracy of both tumors and OARs. This precision facilitates more exact dose calculations, reducing radiation exposure to healthy tissues and improving overall treatment outcomes. Moreover, PCCT offers more precise estimations of proton SPR, crucial for effective proton therapy. By minimizing uncertainties in SPR estimates, PCCT ensures the accurate delivery of the prescribed dose, reducing the risks of under or overdosing.

PCCT has the potential to transform internal radiotherapy techniques such as brachytherapy and RPT, through its advanced imaging features. In brachytherapy, the reduction of metal artifacts by PCCT can improve the visibility around implanted radioactive seeds, leading to more accurate treatment planning. In RPT, enhanced tracking and quantification of radiopharmaceuticals enable more precise dosimetry calculations, ensuring optimal radiation dose delivery to tumors while minimizing exposure to healthy tissues.

The evolving practice of ART stands to benefit immensely from PCCT's superior imaging capabilities. High-resolution images are crucial for detecting small anatomical changes during treatment, vital for accurate deformable registration and real-time treatment adjustments. Such precision ensures that radiation doses remain accurately targeted at the tumor, improving the effectiveness of ART and enhancing patient outcomes.

Another potential of PCCT in clinical applications has been illustrated by Bader et al. (Bader *et al* 2020), who designed a photon-counting cone-beam CT (PCCBCT) system with a small detector area for enhanced imaging quality. Their benchtop system utilized a step-and-shoot acquisition, a user-friendly control interface, and iterative optimization for better geometrical parameter estimation. This lab-based success underscores PCCT's potential in clinical settings requiring higher spatial resolution and advanced spectral imaging. Future applications could include optimizing online ART workflows, where precise imaging is essential for deformable registration and motion tracking.

Additionally, radiomics and deep learning, two rapidly advancing fields, could also significantly benefit from PCCT's advanced imaging features. Radiomics involves extracting quantitative imaging features to build predictive models for treatment outcomes, while deep learning algorithms automate image segmentation and tumor response evaluation. The multi-energy, high-resolution images provided by PCCT enrich these models, enhancing their predictive accuracy and supporting more personalized treatment plans. Compared to conventional CT, PCCT-based radiomic features and deep learning models trained on PCCT data have shown to improve accuracy in tumor stratification and treatment response evaluation.

Lastly, quantitative imaging facilitated by PCCT is pivotal in evaluating tumor responses and guiding adaptive treatment strategies. Detailed data on tumor morphology, density, and perfusion provided by PCCT enable clinicians to monitor and adapt treatment plans effectively, optimizing therapeutic efficacy while minimizing the risks associated with over or undertreatment.

**10.2 Comparison with other CT Modalities**

As shown in Figure 1 and supported by the results discussed, limited studies exist regarding the direct application of PCCT in radiotherapy. To provide a more comprehensive review, we considered studies related to spectral CTs such as DECT, DLCT, and MECT. The comparisons and findings in Sections 3 and 4 demonstrate the clear advantage PCCT holds over these spectral CT methods in terms of spatial resolution, contrast resolution, and overall imaging quality. Radiotherapy workflows that benefit from the enhancements of spectral CTs are likely to experience even greater improvements with PCCT, given its superior spatial and contrast resolution.

Furthermore, the distribution of studies involving PCCT, and other CT types (Table 10) discussed in this paper illustrates the current focus areas of research in RT with PCCT, highlighting the need for more in-depth investigation in critical areas such as image segmentation, proton SPR estimation, and tumor response assessment. These studies reveal that while PCCT is advancing in specific areas like dose reduction, more comprehensive reviews and applications, particularly in proton SPR estimation, are present in studies focusing on other CT modalities.

**Table 10**. Comparison of PCCT and other CT modalities in radiotherapy studies.

| Category | PCCT | Other CT | Comparison |
|---|---|---|---|
| Reviews | 7 | 2 | 0 |
| Segmentation | 1 | 4 | 2 |
| Dose Calculation | 1 | 3 | 0 |
| Proton SPR | 3 | 17 | 4 |
| MAR | 2 | 2 | 0 |
| ART | 1 | 0 | 0 |
| Tumor Response | 2 | 11 | 1 |
| Radiomics and DL | 4 | 6 | 3 |
| Internal Radiotherapy | 0 | 3 | 0 |
| CT Dose Reduction | 10 | 3 | 3 |

### 10.3 Radiation Dose Reduction Benefits

One of PCCT's most notable advantages is its ability to reduce radiation dose by 30-60%, while maintaining or even improving image quality. This dose reduction is particularly valuable in radiotherapy, where repeated imaging is required for treatment monitoring. Studies such as those by Donuru et al (Donuru *et al* 2023) and Jungblut et al (Jungblut *et al* 2022) have consistently shown that PCCT achieves significant dose reductions without compromising diagnostic accuracy.

Pediatric patients, who are more sensitive to radiation-induced risks (Brenner *et al* 2001, Kleinerman 2006, Kutanzi *et al* 2016), benefit significantly from lower exposure during repeated imaging sessions. In proton therapy, where precise SPR estimation is critical, PCCT minimizes dose uncertainties, ensuring safer and more effective treatments for children.

Elderly patients and those with co-morbidities also benefit from PCCT's enhanced MAR capabilities, improving imaging quality in the presence of implants. Additionally, patients requiring long-term follow-up after cancer therapy can undergo repeated imaging with lower cumulative radiation doses (Rubino *et al* 2003, Smith-Bindman 2009), reducing potential late effects and improving overall care. This highlights PCCT's potential to address the unique needs of diverse patient populations.

**10.4 Cost Effectiveness of PCCT Systems**

The high cost of PCCT systems remains a significant challenge to their broader adoption in clinical and radiotherapy settings. This expense is driven by the sophisticated detector materials, such as CdTe and CZT, which require high purity and complex manufacturing processes to achieve the precision necessary for photon-counting detectors. In addition, PCCT systems often necessitate multiple generators and X-ray tubes, further contributing to their overall cost (Tortora *et al* 2022, Zanon *et al* 2023, Meloni *et al* 2023a). While this upfront investment may be substantial, PCCT offers opportunities for long-term cost-effectiveness.

One promising approach involves leveraging the advanced imaging capabilities of PCCT to reduce the need for additional scans. In radiotherapy, where contrast-enhanced imaging is crucial for tumor contouring or assessing early treatment response, PCCT's enhanced contrast and spatial resolution could potentially eliminate the need for additional contrast-enhanced scans, thereby saving on operational costs. Moreover, as detector technologies evolve, alternative materials like silicon photomultipliers (SiPMs) are being explored, which may provide comparable performance at reduced costs (van der Sar *et al* 2021).

Another consideration is the potential to extract more diagnostic information from traditionally ancillary scans, such as scout scans. Research has shown that PCCT scout images can accurately measure biomarkers like bone mineral density (BMD) without additional radiation or cost, showcasing the "always on" advantage of PCCT systems (Pourmorteza 2021). This dual utility could offset some of the initial costs by enabling multitasking capabilities within the same imaging workflow.

While the current costs of PCCT systems present barriers to adoption, ongoing advancements in detector design, calibration algorithms, and AI-driven optimization have the potential to make this transformative technology more accessible and economically feasible for widespread use in radiotherapy and beyond.

**10.5 Challenges and Limitations**

Despite its remarkable benefits, PCCT faces certain limitations that hinder its broader clinical adoption. While PCCT has shown significant promise in diagnostic imaging, its application in radiotherapy remains underexplored. The higher cost of PCCT systems poses a significant challenge. Many medical institutions, particularly in low-resource settings, may struggle to afford this technology, limiting its global accessibility. Moreover, PCCT-specific software packages for radiotherapy are still in early development. Essential tools like motion tracking, quality assurance (QA), and advanced image guidance—critical for radiotherapy workflows—are not fully optimized. This limits the seamless integration of PCCT into clinical practice.

Technical constraints such as detector size and scan time also need to be addressed to ensure seamless integration into clinical workflows. The smaller detector size of many PCCT systems limits the field of view (Flohr *et al* 2020b, Meloni *et al* 2023b, Zhan *et al* 2023), which may pose challenges in imaging large anatomical areas or accounting for patient motion during acquisition. Similarly, the increased data requirements for photon-counting detectors can result in longer scan times (Wu *et al* 2023, Dane *et al* 2024a), potentially impacting workflow efficiency in busy radiotherapy departments. Continued advancements in detector technology and acquisition protocols are essential to overcome these barriers and unlock PCCT's full potential in radiotherapy. While PCCT offers significant potential for radiation dose reduction, certain high-resolution or multi-energy imaging applications may necessitate increased photon flux to maintain image quality, potentially leading to higher radiation doses (Flohr *et al* 2023, Sartoretti *et al* 2023). This underscores the importance of balancing imaging performance with patient safety, particularly in applications demanding ultra-high resolution or complex spectral imaging.

While PCCT technology demonstrates clear advantages in diagnostic imaging, its clinical integration into radiotherapy settings is hindered by lack of clinical trials. A review of existing studies on clinicaltrials.gov revealed several trials evaluating PCCT compared to conventional EID-CT systems for diagnostic purposes. For instance, NCT05838482 (recruiting) investigates PCCT's utility in reducing radiation dose and enhancing image quality metrics such as spatial resolution and image contrast, supporting its potential in general clinical applications. Similarly, NCT06281808 (active, not recruiting) compares PCCT and EID-CT for musculoskeletal imaging, emphasizing higher spatial resolution and dose efficiency.

Completed trials, like NCT03878134, have demonstrated PCCT's improved imaging quality for routine diagnostic scans in select patient groups. Meanwhile, studies such as NCT05240807 focus on leveraging PCCT's advanced capabilities for coronary artery disease diagnosis, exploring its spatial and temporal resolution benefits. Upcoming efforts, including NCT06691659, aim to evaluate a next-generation PCCT system's performance across various diagnostic applications, including abdominal, cardiothoracic, and neuroimaging.

However, none of these studies directly address PCCT's efficacy in radiotherapy workflows. The absence of trials evaluating critical radiotherapy parameters, such as tumor delineation accuracy, proton SPR estimation, and treatment response monitoring, highlights a significant research gap. Designing randomized controlled trials to assess these applications is imperative. Such efforts would provide robust data to validate PCCT's utility in radiotherapy, establish its safety and efficacy, and guide its integration into clinical workflows. By bridging this gap, PCCT could fulfill its transformative potential in precision oncology.

While there are limited long-term follow-up studies explicitly assessing tumor control and normal tissue toxicity for treatments planned using PCCT, its superior imaging capabilities are anticipated to contribute to improved clinical outcomes. By enhancing tumor delineation accuracy and reducing uncertainties in proton SPR, PCCT can improve dose conformity and spare healthy tissues more effectively than conventional CT methods. These advancements suggest potential for reduced long-term side effects and better tumor control rates. However, prospective trials and longitudinal studies are essential to validate these theoretical benefits and establish PCCT's impact on patient outcomes in radiotherapy.

### 10.6 Future Directions and Innovative Solutions

To address these limitations of PCCT, innovative solutions are necessary. Artificial intelligence (AI) offers a promising approach to reducing costs and increasing global accessibility. AI algorithms, including diffusion models, CycleGANs, and other generative models, could be trained to synthesize PCCT images from conventional EICT data. This would reduce reliance on expensive PCDs by providing institutions with EICT systems the ability to access PCCT-quality images. Additionally, AI could streamline processes like image registration, dose recalculation, and segmentation, reducing operational costs and improving efficiency in radiotherapy departments. AI-driven adaptive radiotherapy, utilizing real-time PCCT images, could further enhance treatment precision by adjusting radiation dose based on changes in patient anatomy.

A critical factor in making PCCT more practical for radiotherapy is the development of specialized software by vendors. Currently, most PCCT software is designed for radiology, leaving radiation oncology departments without essential tools for QA, motion tracking, and image guidance. Vendors must collaborate with radiation oncology departments to create software that fully integrates PCCT's advanced features, including multi-energy image reconstruction, metal artifact reduction, and advanced dose calculation algorithms. These tools would enhance the precision of radiotherapy treatments, enabling real-time adjustments and improving patient outcomes.

Exploring alternative detector technologies could also help reduce the cost of PCCT systems. Silicon photomultiplier (SiPM)-based detectors, as explored by (van der Sar *et al* 2021) offer a promising alternative to CdTe and CZT detectors. These detectors can achieve similar energy resolution and rate capability while potentially reducing system costs, making PCCT more accessible for radiotherapy applications.

Monte Carlo simulations offer another solution for optimizing PCCT-based radiotherapy protocols. As discussed by Wang et al (Wang *et al* 2020, 2024) and Li et al (Li *et al* 2024a), these simulations provide detailed modeling of photon transport and tissue interactions, which can be used to validate new PCCT protocols before clinical implementation. Monte Carlo studies play a crucial role in developing PCCT for complex treatment scenarios like proton therapy, where accurate dose calculations are critical.

In immunotherapy, PCCT's ability to provide detailed multi-energy imaging enhances early detection of treatment responses, allowing clinicians to adapt therapies based on individual patient dynamics.

By addressing these challenges and leveraging innovations such as AI-driven optimization, advanced detector technologies, and targeted clinical research, PCCT is poised not only to revolutionize radiotherapy workflows but also to redefine the standards of precision and efficacy in cancer treatment, ensuring better outcomes for patients across diverse populations.

## 11.Conclusion

In conclusion, PCCT offers numerous advantages over conventional EICT systems, but addressing its limitations is essential for broader adoption in radiotherapy. AI-driven solutions, specialized software development, alternative detector technologies, and Monte Carlo simulations all have the potential to make PCCT a more accessible and integrated tool in cancer treatment. PCCT's ability to revolutionize radiotherapy—from enhancing tumor targeting to enabling adaptive treatments—positions it as a key technology in the future of cancer diagnosis and treatment.

## Acknowledgements

This research is supported in part by the National Institutes of Health under Award Numbers R01CA272991, R01DE033512, U54CA274513, and R01EB032680.